\documentclass[aps,prx,twocolumn,a4paper,superscriptaddress,10pt]{revtex4-2} 
\usepackage{etoolbox}
\apptocmd{\sloppy}{\hbadness 20000\relax}{}{}

\usepackage[english]{babel}
\usepackage{graphicx}
\usepackage{hyperref}
\hypersetup{
    colorlinks,
    linkcolor={red!50!black},
    citecolor={blue!50!black},
    urlcolor={blue!80!black}}
\usepackage{cancel}
\usepackage{amsmath}
\usepackage{amsfonts}
\usepackage{amssymb}
\usepackage{bm}

\usepackage{physics}
\usepackage{relsize}
\usepackage{mathtools}
\usepackage{indentfirst}
\usepackage{pstricks}
\usepackage{tikz}
\tikzset{>=latex}
\usetikzlibrary{patterns}
\usepackage[overload]{empheq}
\usepackage{xcolor}
\usepackage{natbib}
\usepackage{dsfont}
\usepackage{microtype}
\usepackage{mathrsfs}

\usepackage[caption=false]{subfig}

\makeatletter
\renewcommand{\fnum@figure}{FIG. \thefigure}
\makeatother

\usepackage{stmaryrd}
\SetSymbolFont{stmry}{bold}{U}{stmry}{m}{n}
\usepackage{wrapfig}

\usepackage[version=4]{mhchem}

\usetikzlibrary{trees,arrows.meta}
\usetikzlibrary{arrows}

\makeatletter
\newcounter{savesection}
\newcounter{apdxsection}
\renewcommand\appendix{\par
  \setcounter{savesection}{\value{section}}%
  \setcounter{section}{\value{apdxsection}}%
  \setcounter{subsection}{0}%
  \gdef\thesection{\@Alph\c@section}}
\newcommand\unappendix{\par
  \setcounter{apdxsection}{\value{section}}%
  \setcounter{section}{\value{savesection}}%
  \setcounter{subsection}{0}%
  \gdef\thesection{\@arabic\c@section}}
\makeatother

\begin{document}
\title{Resolving Speed and Encoding Bottlenecks in Fast Heteromeric Self-Assembly}
\author{Félix Benoist}
\affiliation{Gulbenkian Institute of Molecular Medicine, Oeiras, Portugal}

\author{Pablo Sartori}
\email{pablo.sartori@gimm.pt}
\affiliation{Gulbenkian Institute of Molecular Medicine, Oeiras, Portugal}

\begin{abstract}
The cytoplasm is a heterogeneous mixture containing many types of proteins that self-assemble into a wide variety of complexes. The accuracy and speed of cytoplasmic self-assembly is astonishing because it involves the correct identification of components shared among different structures, despite pervasive thermal fluctuations. Typical toy models of self-assembly are based on the specificity of binding energies among components. 
However, kinetics plays a key role in biological self-assembly, often catalyzed by a plethora of assembly factors. Building on this observation, we extend a previous heteropolymer growth model to describe the retrieval of two-dimensional structures via quasi-2D growth. We find that the self-assembly of structures in this model is subject to strong speed and encoding bottlenecks. Moreover, we show that these bottlenecks can be suppressed by increasing the connectivity of a small fraction of components. This mechanism of kinetically controlling a small number of critical binding events provides a simple explanation for the timely assembly of large protein, and suggests a unifying principle for the role of assembly factors.
\end{abstract}
\maketitle

\section{Introduction}\label{sec.intro}
Many proteins assemble into heteromeric complexes, \emph{i.e.} complexes composed of multiple protein species, to carry out essential cellular functions. During assembly, protein complexes must accurately identify their specific components from a crowded mixture of thousands of species. Assembly must not only be precise but also rapid, so that cellular functions proceed in a timely manner. Two remarkable examples are the spliceosomal complexes and the ribosomal subunits, each composed of dozens of proteins~\cite{Khatter15, Wahl09}, with ribosomal subunits assembling in under 2 minutes~\cite{Chen12}. Advances in high-throughput cryoEM have steadily expanded our knowledge of the molecular and structural details of such complexes. Yet, the generic physical principles that enable high assembly speed and low error rates to coexist in heterogeneous cellular environments remain elusive.

\begin{figure*}[!t]
    \includegraphics[width=\linewidth]{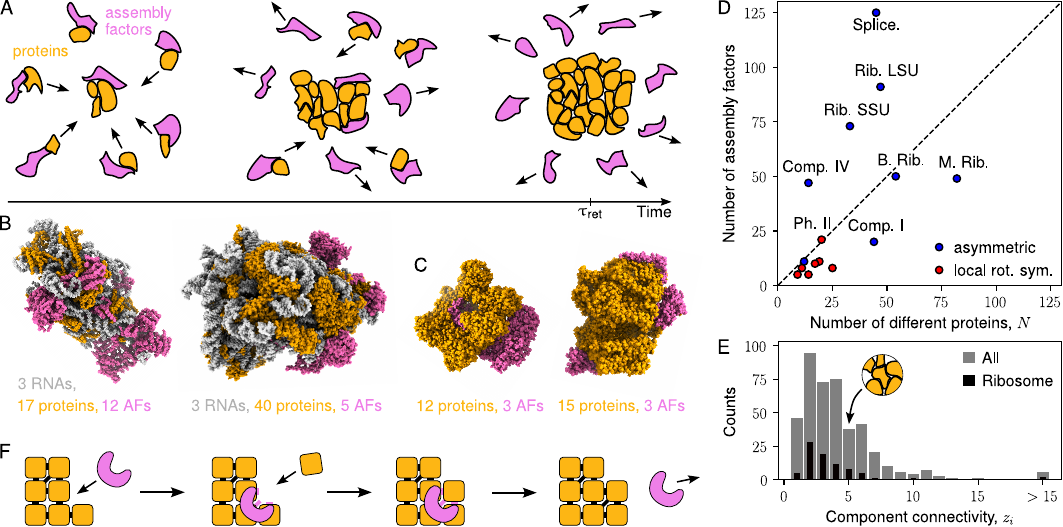}
   \caption{{\it Formation of large heteromeric complexes rely on many assembly factors.} {\bf A.} Schematic representation of the self-assembly of a protein complex. The self-assembly of protein subunits is driven by assembly factors, that have the ultimate role of favoring speed and accuracy of complex formation. The complete protein assembly contains no assembly factor.
   {\bf B.} Structural snapshots of assembly intermediates of the yeast ribosome large subunit. As we can see, assembly factors transiently bound to the RNA-protein assembly. Protein data bank IDs: 6EM3, 6LSR~\cite{Berman00}.
   {\bf C.} Intermediates of the human proteasome (8QZ9) and of the bacterial photosystem II (7NPQ). 
   {\bf D.} In purely asymmetric complexes, numbers of assembly factors roughly increase with numbers of protein subunits. By contrast, complexes with a local rotational symmetry require fewer assembly factors. Data detailed in~\cite{SM} Sec.~\ref{sec.SI-Data} and Refs.~\cite{Wahl09,Dorner23,Woolford13,Lavdovskaia24,Vercellino22, Heinz16,Yang15,Wild12,Ruhle15,Altegoer15,Armitage20,Rousseau18} therein.
   {\bf E.} The connectivity of components in an assembly is highly variable. Highly-connected components correspond to RNAs. Detailed plot in Fig.~\ref{fig:con_detailed}. 
   {\bf F.} Our kinetic encoding model consists in the acceleration of correct binding events via kinetic cooperativity and catalyzing assembly factors.}
\label{fig:prot_comp}
\end{figure*}

What are the key biological ingredients of cellular self-assembly? We identify two such ingredients: assembly factors and assembly geometry, which we describe below. Assembly factors are proteins that transiently assist the growth of complexes~\cite{Dorner23,Wahl09,Vercellino22}, akin to chaperones in protein folding; schematic in Fig.~\ref{fig:prot_comp}A. Assembly factors bind at specific locations and times during the growth of protein complexes; examples in Figs.~\ref{fig:prot_comp}B and C. Moreover, their number is often comparable to that of the proteins in the complex itself; see Fig.~\ref{fig:prot_comp}D. Their molecular functions are diverse: some block premature binding or catalyze specific events~\cite{Davis17,Rousseau18,Altegoer15,Dorner23}, and yet others act as proofreading enzymes~\cite{Dorner23,Seffouh24,Wahl09}. Overarching this diversity is the observation that depletion of assembly factors slows down self-assembly and leads to the accumulation of incomplete intermediates~\cite{Seffouh24,Davis17,Mulder10,Sheng23}. Another common feature is that assembly factors are dissipative, hydrolyzing ATP or GTP to perform their function~\cite{Davis17,Dorner23,Altegoer15,Wild12}. Thus, assembly factors are a key ingredient in protein complex assembly, ensuring both speed and accuracy by strongly driving specific binding events.

A second fundamental ingredient is the diversity of protein shapes and connectivities within heteromeric complexes. Even at a coarse grained level, some proteins (or RNA components) are large and elongated, while others are small and globular~\cite{Erickson09,Reuveni17}. Consequently, their connectivity varies widely: Some proteins bind to many other proteins in the final complex, while others bind to few; see Fig.~\ref{fig:prot_comp}E for examples across complexes. This structural diversity has important kinetic consequences. In particular, the kinetic cooperativity of neighboring proteins is known to largely accelerate self-assembly~\cite{Bunner10,Williamson08}. We thus expect that highly connected proteins can be kinetically discriminated and quickly assemble without errors. Overall, unlike large homomeric complexes such as actin filaments~\cite{Chhabra07,vonderEcken15}, microtubules~\cite{Fellous77} or viral capsids~\cite{Mateu13} which are highly symmetric and require few assembly factors, large heteromeric complexes feature far more intricate geometries and involve many catalyzing assembly factors.

In this paper, we develop a toy model of heteromeric self-assembly to investigate the role of the two ingredients above. Our goal is not to undertake detailed molecular dynamics of heteromeric self-assembly~\cite{Kurisaki22,Perilla15}; but rather to design simplified models that illuminate fundamental principles of nonequilibrium self-assembly and geometry, and their implications in biological and synthetic systems~\cite{Gartner24,Lenz17,Manoharan15,Ravasio24}. Building on previous work in self-assembly~\cite{Gartner24,Murugan_PNAS15,Sartori20,Benoist25_SA} and nonequilibrium information processing~\cite{Dill93,Bennett79,Sartori13,Tsai06,Benoist25_Hop}, we explore how kinetic discrimination and variable connectivity enable fast and accurate assembly in the presence of confounding interactions. 

The paper is structured as follows. In Sec.~\ref{sec.model}, we introduce a minimal lattice model for far-from-equilibrium heteromeric self-assembly. In Sec.~\ref{sec.speed_bot}, we study the assembly of a single square target with fixed connectivity using Monte Carlo simulations and identify speed bottlenecks. In Sec.~\ref{sec.enc_bot}, we extend the model by encoding an additional structure with the same components and show that the speed bottlenecks also act as encoding bottlenecks. In Sec.~\ref{sec.theory}, we derive conditions for rapid assembly of multiple structures and propose alleviating bottlenecks by increasing the connectivity of a small subset of components. In Sec.~\ref{sec.sol}, we numerically validate these analytical predictions. Finally, in Sec.~\ref{sec.combi}, we show that increased connectivity yields combinatorial scaling of the encoding capacity.

\section{Model setup}\label{sec.model}
Within our framework, a set of square tiles with varying connectivity can reversibly bind at specific locations on the boundary of a growing structure, aided by catalyzing assembly factors; see Fig.~\ref{fig:prot_comp}F. We abstract the action of assembly factors and the kinetic cooperativity discussed in the introduction by accelerating binding events that lead toward the formation of a target two-dimensional structure representing a protein complex. The corresponding unbinding events are likewise accelerated, so that the target structure is encoded purely kinetically. To capture the promiscuous nature of proteins~\cite{Gavin06,Kuhner09}, which may lead to erroneous bonds between monomers not in contact in the final structure, we allow multiple target structures to be encoded simultaneously from the same components. In this setting, monomer promiscuity increases uniformly with the number of encoded structures. Our model thus constitutes an abstraction of protein complex assembly, which we now describe in greater mathematical detail.

\begin{figure}[!t]
	\centering
	\includegraphics[width=\linewidth]{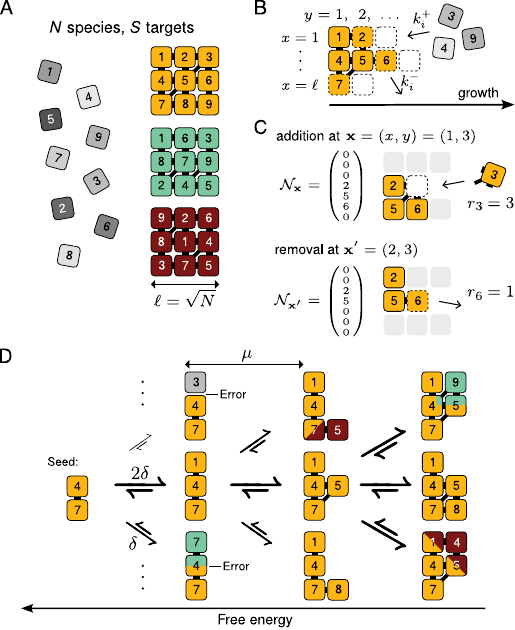}   
	\caption{{\it Kinetic encoding of multiple target structures.} 
	{\bf A.} A mixture of $N=9$ component species can assemble into $S=3$ structures of size $N=\ell\times\ell$. Components bind via the black rectangles. Vertical periodic boundaries imply that, \emph{e.g.} in the yellow target, components 1, 2 and 3 also bind to 7, 8 and 9.
    {\bf B.} Growth occurs by adding/removing monomers at the boundary locations $\mathbf x=(x,y)$ (dashed contour). 
    {\bf C.} Addition ($k_i^+$) and removal ($k_i^-$) rates depend on the number of correct partners ($r_i$) in the neighborhood $\mathcal N_{\bf x}$ (0 denotes empty sites). Here, component 3 binds to the 3 neighbors as in the yellow target, while component 6 binds to only one of its two neighbors. Note that bonds serve to accelerate binding, rather than to stabilize it. 
    {\bf D.} Example self-assembly pathway of the yellow structure from a small seed. Each monomer binding decreases free energy via $\mu$. The acceleration of correct binding/unbinding events via $\delta$ [Eq.~\eqref{eq:rates}] is sketched as larger arrows. Binding of component 3 at the first assembly step is a thermal error, due to thermal fluctuations; whereas binding of component 7 is a combinatorial error, due to component reuse across targets.} 
\label{fig:sketch}
\end{figure}

We consider a set of $N\gg1$ component species labeled $i=1,\ldots,N$; see Fig.~\ref{fig:sketch}A. Each component corresponds to a tile with a fixed orientation on a square lattice, that can bind with its neighbors. All tiles have the usual nearest-neighbor binding sites (north, n; south, s; east, e; west, w), and some tiles may have additional next-nearest-neighbor binding sites (north-east, ne; north-west, nw; south-east, se; south-west, sw). 
The components are held at fixed chemical potentials, $\mu_i=\mu$, and may self-assemble into $S\ge1$ heteromeric square target structures of side length $\ell=\sqrt N$. Each target corresponds to a reshuffling of all $N$ species, so $S$ quantifies the degree of promiscuity. Crucially, reuse of components across different structures compromises the reliability of self-assembly, a phenomenon previously described as multifarious~\cite{Murugan_PNAS15, Sartori20,Osat23}.

To study retrieval of a structure, we model growth from a nucleation seed as a boundary-driven process, in which components bind/unbind at lattice locations $\mathbf x = (x, y)$ of a single boundary of the assembly~\cite{Whitelam14,Nguyen16}; see Fig.~\ref{fig:sketch}B as well as~\cite{SM} Sec.~\ref{sec.SI-Methods}. The perpendicular lattice direction has periodic boundary conditions connecting the lattice locations $x=\ell$ to $x=1$. This setup corresponds to a quasi-2D assembly process. The rates $k_i^\pm$ for adding and removing component $i$ at $\mathbf x$ depend on the identity of the components in the neighborhood $\mathcal N_{\bf x}$, which we write as a 7-dimensional vector: $\mathcal N_{\bf x}=(i_{\rm ne}, i_{\rm n},\dots,i_{\rm se})$; see Fig.~\ref{fig:sketch}C for examples. (The eastern neighbor is not taken into account since components with an eastern neighbor are not part of the boundary.)
To enable fast and reliable self-assembly of multiple targets, we encode structures in the binding kinetics, instead of the energetics. Following earlier work on polymerization~\cite{Benoist25_SA,Sartori13,Bennett79}, we define
\begin{equation}\label{eq:rates}
\begin{aligned}
	k_i^+(\mathcal N_{\bf x})&=\exp(r_i(\mathcal{N}_{\bf x})\delta) \quad,\\
	k_i^-(\mathcal N_{\bf x})&=\exp(r_i(\mathcal{N}_{\bf x})\delta-\mu)\quad.
\end{aligned}
\end{equation}
In this expression, $\delta$ is a kinetic discrimination parameter that quantifies differences in kinetic barriers, and $r_i(\mathcal{N}_{\bf x})$ counts the number of neighbors in $\mathcal N_{\bf x}$ that also neighbor $i$ at the corresponding relative locations in any target structure (not necessarily the one being assembled); see~\cite{SM} for explicit expression. Thus, binding of $i$ at location $\bf x$ creates $r_i(\mathcal{N}_{\bf x})$ bonds. Note that, since $k_i^+/k_i^-=\exp(\mu)$, the free-energy change due to growth contains no structural information, unlike in Refs.~\cite{Sartori20, Murugan_PNAS15, Chen25, Bisker18, Bupathy22, Holmes-Cerfon25}. Instead, Eq.~\eqref{eq:rates} accelerates addition/removal events that are consistent with target neighborhoods, so the structures are kinetically encoded; see schematics in Fig.~\ref{fig:sketch}D. In~\cite{SM} Sec.~\ref{sec.SI-coop-AF}, we show how Eq.~\eqref{eq:rates} can be interpreted within a simple framework combining kinetic cooperativity and assembly factor catalysis.

In the case where components participate in multiple targets ($S>1$), their connectivity may differ from one structure to another. We thus define $z_i$ as the connectivity of component $i$ within the seeded structure. Components with higher connectivity ($z_i$ large) can achieve larger $r_i(\mathcal{N}_{\bf x})$, enabling faster binding and stronger discrimination than weakly connected component ($z_i$ small). For instance, in Fig.~\ref{fig:sketch}, component $i=5$ has $z_5=6$ in the yellow target due to two diagonal bonds. As a consequence, its addition rate at the correct location ${\bf x}=(2,2)$ in the growing yellow structure of Fig.~\ref{fig:sketch}D is $k_5^+(\mathcal N_{\bf x})=\exp(2\delta)$, compared to $k_5^+(\mathcal N_{\bf x'})=\exp(\delta)$ at a location ${\bf x}'=(3,2)$ consistent with the red structure. 

We track assembly size by counting the number of monomers on the lattice in units of $N$, and we define two measures of accuracy of tiles, a global and a local one. Global accuracy is measured as the fraction of correctly placed tiles relative to the seeded target within the square $\mathbf x=(1,1)$ to $(\ell,\ell)$ [Fig.~\ref{fig:sketch}B]. A tile contributes to the global accuracy of the seeded structure regardless of its local neighborhood. By contrast, local accuracy is defined through bonds: a tile is locally accurate if $r_i(\mathcal{N}_{\bf x})\ge1$. This measure is used to identify chimeric structures. We moreover characterize assembly dynamics via two characteristic timescales. The retrieval time $\tau_{\rm ret}$ corresponds to the time it takes to retrieve a target structure, \textit{i.e.} the time for the global accuracy to reach unity. Likewise, we define the lifetime $\tau_{\rm life}$ as the time it takes to add one extra monomer to the completed structure.

In the following, we analyze retrieval of kinetically-encoded structures under the rates of Eq.~\eqref{eq:rates} for mixtures of components with varying connectivity.

\section{Speed bottlenecks} \label{sec.speed_bot}
\begin{figure}[!t] 
   \centering
   \includegraphics[width=\linewidth]{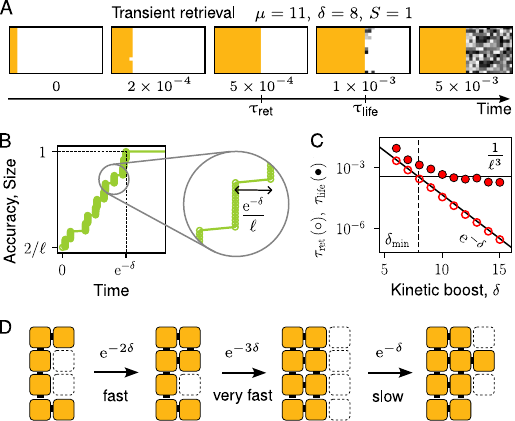}
   \caption{{\it Nearest-neighbor case yields slow retrieval.} 
   {\bf A.} Assembly snapshots showing retrieval of a single square target of side length $\ell=14$ tiles from a two-layer nucleation seed. Assembly is allowed to grow beyond the target size $\ell$ up to $2\ell$. White denotes empty locations, yellow shows correctly placed tiles, and shades of gray denote incorrect tiles of different species (actually corresponding to chimeras of the target). Other growth regimes are described in~\cite{SM} Sec.~\ref{sec.SI-Growth}.  
   {\bf B.} The global retrieval accuracy, \textit{i.e.} the fraction of correctly placed tiles (line), follows the normalized assembly size (circles) over time and exhibits speed bottlenecks.
   {\bf C.} The target retrieval time ($\tau_{\rm ret}$) from Eq.~\eqref{eq:t_ret} separates from the target lifetime ($\tau_{\rm life}$) for $\delta>\delta_{\rm min}$ [Eq.~\eqref{eq:d_min}].
   {\bf D.} In the irreversible, high-discrimination regime, each addition event maximizes the number of bonds. The mean waiting time is dominated by the inverse of the addition rate [Eq.~\eqref{eq:rates}], so additions with few bonds slow down the assembly.}
\label{fig:Z4}
\end{figure}

As a starting point, we consider a mixture of $N$ components, whose kinetics encode a single target structure, $S=1$, with bulk connectivity $z=4$. (While bulk components have $z_i=4$ encoded bonds, boundary components only have $z_i=3$.) This corresponds to a simple model of heteromeric 2D self-assembly~\cite{Sartori20,Murugan_PNAS15,Bisker18, Gartner24,Holmes-Cerfon25,Osat23,Bupathy22,Ravasio24}. Based on previous work on kinetic encoding~\cite{Sartori13,Benoist25_SA,Benoist25_Hop}, successful self-assembly is expected when the kinetics are both highly irreversible, large $\mu$, and highly discriminatory, large $\delta$. Figure~\ref{fig:Z4}A shows snapshots of such a growing assembly from a small nucleation seed. In this regime, the target structure (yellow) is retrieved within a time $\tau_{\rm ret}$, but because it is not a free-energy minimum (\textit{i.e.} not thermodynamically stable), incorrect tiles are eventually incorporated after a lifetime $\tau_{\rm life}$. 

A closer analysis of the retrieval dynamics reveals that $\tau_{\rm ret}$ is not governed by a typical timescale but instead by a few rare, slow events, see Fig.~\ref{fig:Z4}B, unlike in the case of heteropolymer growth~\cite{Benoist25_SA}. Specifically, these bottleneck events have a characteristic time $\exp(-\delta)$, so that the retrieval time scales as $\tau_{\rm ret}\approx\exp(-\delta)$ [Fig.~\ref{fig:Z4}C]. By contrast, the target lifetime $\tau_{\rm life}$ is asymptotically independent of $\delta$, ensuring kinetic stability of the structure. Yet, a much larger timescale separation is prevented due to the few events that act as speed bottlenecks. 

The origin of the bottlenecks can be understood intuitively as follows (a formal derivation is carried out in Sec.~\ref{sec.theory}). In the highly irreversible regime, assembly dynamics are governed by the addition rates $k_i^+$. For any boundary configuration, the most probable next additions are those with the largest $k_i^+$. Assuming error-free growth, addition events fall into three classes: $\order{N}$ addition events correspond to lateral layer growth, with $k_i^+=\exp(2\delta)$; $\mathcal O(\sqrt{N})$ events close a layer, with $k_i^+=\exp(3\delta)$; and $\mathcal O(\sqrt{N})$ events nucleate a new layer, with $k_i^+=\exp(\delta)$. As illustrated in Fig.~\ref{fig:Z4}D, the slowest of these, \textit{i.e.} nucleation of new layers, sets the overall scaling of the retrieval time. Thus, although most growth events are fast, lasting a mean time $\exp(-2\delta)$, the rare bottleneck events dominate the assembly time, leading to $\tau_{\rm ret}\approx\exp(-\delta)$.

\section{Encoding bottlenecks}\label{sec.enc_bot}

\begin{figure}[!t] 
   \centering
   \includegraphics[width=\linewidth]{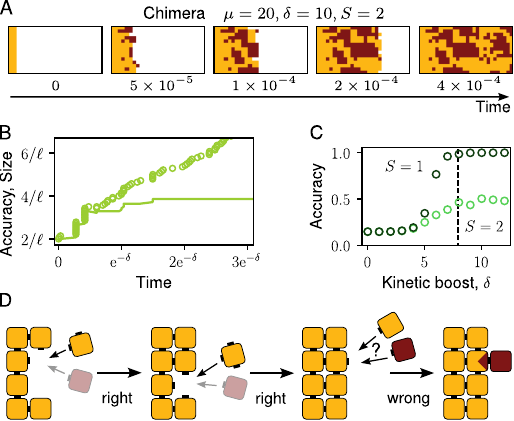}
   \caption{{\it Speed bottlenecks are also encoding bottlenecks.} 
   {\bf A.} Assembly snapshots for $S=2$ arbitrary targets (yellow and red) show the formation of a chimeric structure, detailed description in~\cite{SM} Sec.~\ref{sec.SI-Growth}. Here, tiles are colored according to the target with which they share the most correct neighbors, irrespective of global accuracy.
   {\bf B.} Errors make the global retrieval accuracy (line) depart from the normalized assembly size (circles). 
   {\bf C.} For $S=2$ targets, the maximal accuracy at large $\delta$ is around 50\%.
   {\bf D.} Since each component has two different partners, kinetic discrimination is impossible at addition events with only one bond.}    
 \label{fig:Z4_S=2}
\end{figure}

So far, we have shown that for large $\mu$ and $\delta$ a single target structure can be reliably assembled, albeit with speed bottlenecks. We now turn to the case in which the same mixture encodes two distinct assemblies, $S=2$. (Bulk components thus have a set of $z_i=4$ partners in each structure.) Figure~\ref{fig:Z4_S=2}A shows that, in this setting, a seed corresponding to a given target (yellow) does not grow reliably. Instead, the resulting structure is a chimera: fragments of the alternative target (red) appear interwoven with those of the seeded target (yellow). Consequently, although the assembly grows over time, retrieval accuracy remains low, around 50\%; see Fig.~\ref{fig:Z4_S=2}B. Importantly, this low fidelity cannot be remedied by increasing the discrimination energy $\delta$; see Fig.~\ref{fig:Z4_S=2}C. The reason is that the errors are not of thermal origin, but of combinatorial origin, as already noted for equilibrium assemblies~\cite{Sartori20,Murugan_PNAS15}.

The emergence of chimeras for $S=2$ is directly linked to the speed bottlenecks identified above. To see this, recall that in the fully irreversible regime the added components are those with maximal rate $k_i^+$. For components of the seeded target (yellow), these rates scale as $\exp(\delta)$, $\exp(2\delta)$, and $\exp(3\delta)$ for nucleating, growing, and closing a layer, respectively. By contrast, erroneous additions from the confounding target (red) almost never benefit from multiple coincident neighbors, and thus their rate scales as $\exp(\delta)$. As a result, at the onset of each new assembly layer, the correct and incorrect components compete on equal footing, leading to a roughly $50\%$ chance of error; see Fig.~\ref{fig:Z4_S=2}D. These rare but slow nucleation events thus have two implications: they limit the overall growth speed and, at the same time, render kinetic discrimination ineffective, producing encoding bottlenecks. 

In the next section, we characterize the general conditions required for both fast and accurate retrieval, and propose a simple change in our model that resolves these dual bottlenecks.

\section{Conditions for fast retrieval}\label{sec.theory}
So far, we have seen that fast retrieval of target structures is possible but limited by encoding and speed bottlenecks; see Figs.~\ref{fig:Z4} and \ref{fig:Z4_S=2}. In this section, we determine the conditions on the chemical potential $\mu$, the discrimination barrier $\delta$, the number of encoded structures $S$, and the assembly size $N$, that allow for fast retrieval of multiple structures. To this end, we unravel the origin of both speed and encoding bottlenecks so that they can be resolved. 

	\subsection{Speed and encoding bottlenecks}
We begin by defining speed and encoding bottlenecks more precisely. Retrieval of kinetically-encoded targets is most effective in the strongly irreversible and highly discriminatory limit, $\mu\to\infty$ and $\delta\to\infty$. In this scenario, the assembly of a structure of size $N$ from a seed of size $N_{\rm seed}$ proceeds through exactly $N_{\rm tot}=N-N_{\rm seed}$ successive addition events, without removals [Eq.~\eqref{eq:rates}]. 

We label these $N_{\rm tot}$ subsequent addition events in order of occurrence by a ``time'' index $t=1,2,\dots,N_{\rm tot}$, distinct from the physical time $\tau$. Each event is characterized by a pair $(i^t, {\bf x}^t)$ indicating the component $i^t$ added at location ${\bf x}^t$. Given that, at time $t-1$, the assembly is characterized by neighborhoods $\mathcal{N}^{t-1}_{\bf x}$, the event at time $t$ is among those that maximize the addition rate:
\begin{align}\label{eq:it,xt}
	(i^t, {\bf x}^t) \in \arg\max_{(i,\bf x)} \left[k_i^+(\mathcal N^{t-1}_{\bf x})\right]\quad.
\end{align}
Typically, multiple components and locations satisfy Eq.~\eqref{eq:it,xt}, and so there are multiple kinetically equivalent events at each time. We further denote the rate corresponding to the addition event $(i^t, {\bf x}^t)$ as $k^t$, and the corresponding number of bonds as $n^t=\ln(k^t)/\delta$; see Eq.~\eqref{eq:rates}. 

Critical events correspond to ``correct'' addition events that are much slower than the rest, potentially as slow as an ``incorrect'' addition. We therefore compute the critical addition rate as the mini-max:
\begin{align}\label{eq:kc,nc}
	k_{\rm c} = \min_{t}k^t = \min_{t} \max_{(i,\bf x)} \left[k_i^+(\mathcal N^{t-1}_{\bf x})\right]\quad,
\end{align}
and correspondingly define $n_{\rm c}=\ln(k_{\rm c})/\delta$. A speed bottleneck occurs when $k_{\rm c}$ corresponds to few events, so that these few slow additions determine the overall retrieval time. An encoding bottleneck occurs when the critical rate $k_{\rm c}$ is equal to the maximal rate of incorrect additions, so that incorrect and correct monomers are indistinguishable.

As an example, consider the case in which structures have a bulk connectivity $z=4$, already discussed before. As the assembly grows from a seed, the number of bonds $n^t$ established by each added component changes over time. As sketched in Fig.~\ref{fig:Z4}D, layer nucleation events create $n^t=1$ bond, layer growth events create $n^t=2$ bonds, and layer completion events create $n^t=3$ bonds. Because there are much fewer layer nucleation events than the total number of events, and because these events are the slowest, they act as speed bottlenecks, with a critical addition rate of $k_{\rm c}=\exp(\delta)$. Furthermore, if $S=2$ structures are encoded, incorrect additions also occur at rate $\exp(\delta)$, and therefore these nucleation events also constitute encoding bottlenecks; see Fig.~\ref{fig:Z4_S=2}D.

	\subsection{Conditions for accuracy at high speed}
To investigate the conditions for accurate high-speed assembly, we focus on a mixture that encodes a single structure, and so $S=1$. In this scenario, only one correct monomer can be added at each boundary location, and so $\mu>0$ is necessary to ensure growth. To determine the minimum discrimination strength necessary for reliable assembly, $\delta_{\rm min}$, we focus on the $N_{\rm c}\le N_{\rm tot}$ critical events with $k^t=k_{\rm c}$ that may act as speed bottlenecks. For large but finite $\delta$, each critical event may correspond either to a correct or an incorrect addition. Correct additions respect condition~\eqref{eq:it,xt} and have an addition rate $k_{\rm c}=\exp(n_{\rm c}\delta)$, while incorrect additions have a rate $\exp(0)=1$. Denoting the multiplicity of correct and incorrect additions at critical events as $\Omega_{\rm C}$ and $\Omega_{\rm I}$, we show in~\cite{SM} Sec.~\ref{sec.SI-Cond} that the minimum discrimination barrier necessary for accurate high-speed retrieval is 
\begin{equation}\label{eq:d_min}
	\delta_{\rm min}= \frac1{n_{\rm c}}\ln\frac{N_{\rm c}\Omega_{\rm I}}{\Omega_{\rm C}}\quad.
\end{equation}
For the simple scenario with connectivity $z=4$ and a square target with side length $\ell=\sqrt N$ [Fig.~\ref{fig:Z4}], we have that addition events that nucleate a new layer create only $n_{\rm c}=1$ bond. At each of these $N_{\rm c}=\ell$ critical events (minus the initial seed length), there is 1 correct monomer and $N-1$ incorrect monomers at each of the $\ell$ locations. This corresponds to a multiplicity of correct additions $\Omega_{\rm C}=\ell$, and a multiplicity of incorrect additions $\Omega_{\rm I}\approx\ell(N-1)\approx\ell^3$. Equation~\eqref{eq:d_min} then yields $\delta_{\rm min}\approx3\ln\ell$. 

Conditions $\mu>0$ and $\delta>\delta_{\rm min}$ result in accurate retrieval of a target structure, as confirmed in Fig.~\ref{fig:Z4}A-B. We can further estimate the time it takes to retrieve a structure from the sum of the mean waiting times of all $N_{\rm c}$ critical events, since these are exponentially slower than the remaining ones. For each critical event, we approximate the mean waiting time as $1/(\Omega_{\rm C} k_{\rm c})$. We therefore have 
\begin{equation}\label{eq:t_ret}
	\tau_{\rm ret}\approx \frac{N_{\rm c}}{\Omega_{\rm C}}\exp(-n_{\rm c}\delta)\quad,
\end{equation}
which reduces to $\tau_{\rm ret}\approx\exp(-\delta)$ for the case $z=4$ shown in Fig.~\ref{fig:Z4}C. By contrast, the lifetime of the target structure, $\tau_{\rm life}$, defined as the time it takes to add a few incorrect monomers, reads $\tau_{\rm life}\approx 1/\Omega_{\rm I}$, because monomers at the target boundary do not have encoded bonds with further monomers. At large $\delta$, these two timescales separate, ensuring kinetic stability of the target. Notice however that in the thermodynamic limit ($N\to\infty$), the minimum discrimination energy diverges logarithmically and the lifetime vanishes. Therefore, kinetic encoding only applies to structures of finite size.

	\subsection{Conditions for combinatorial encoding}
In this subsection, we discuss the conditions for encoding several targets from the same components. When the components of the target being assembled are shared among $S\ge2$ targets, some incorrect additions will also be accelerated via $\delta$. These combinatorial errors arise whenever two components have the same neighbors in different targets. For instance, in Fig.~\ref{fig:sketch}D at time $t=2$, both components 5 and 8 can bind to component 7 with rate $\exp(\delta)$, because they are partners of 7 in the red and yellow structures, respectively. As a result, components 5 and 8 become effectively indistinguishable to the growing assembly. This implies the existence of an encoding capacity $S_{\rm max}$, above which errors cannot be avoided by increasing $\delta$.

To elucidate the origin of these errors, we focus again on the critical events with $k^t=k_{\rm c}$ that can act as encoding bottlenecks. In the simple case $z=4$, each critical addition involves only $n_{\rm c}=1$ bond, which is insufficient to discriminate between components. Indeed, most components have different neighbors in different targets [Fig.~\ref{fig:Z4_S=2}D]. Consequently, already for $S=2$ targets chosen at random, the assembled structure accumulates numerous errors; see Fig.~\ref{fig:Z4_S=2}A. Thus, for $n_{\rm c}=1$, the encoding capacity is limited to $S_{\rm max}=1$.
By contrast, for $n_{\rm c}\ge2$, we find
\begin{equation}\label{eq:Smax_gen}
	S_{\rm max}\sim N^{1-2/n_{\rm c}}\quad
\end{equation}
(\cite{SM} Sec.~\ref{sec.SI-Cond}). The scaling ranges from $\order{1}$ for $n_{\rm c}=2$ to $\order{N}$ in the fully connected case with $n_{\rm c}=N\gg1$.  The corresponding increase in encoding capacity results from a drastic reduction in errors. This is most visible in the low-promiscuity regime, where $S$ exceeds 1 but remains well below $S_{\rm max}$. In that case, we estimate the minimum error fraction at large $\delta$ as 
\begin{equation}\label{eq:error}
	e_{\rm min}\sim N_{\rm c}/N^{n_{\rm c}-1}\quad
\end{equation}
(\cite{SM} Sec.~\ref{sec.SI-Cond}), which decreases rapidly with increasing $n_{\rm c}$.

Next, we examine how reusability of components affects the characteristic timescales of self-assembly. While the retrieval time is still expected to follow Eq.~\eqref{eq:t_ret}, the target lifetime is shortened. This is because components at a target boundary generally belong to the bulk of other targets, and therefore completion of one target accelerates subsequent incorrect additions, yielding $\tau_{\rm life}\lesssim\exp(-\delta)$. More specifically, as the number of structures $S$ grows, they will overlap more, which will decrease $\tau_{\rm life}$ and thus impede timescale separation. 

In summary, the conditions $\mu>0$, $\delta>\delta_{\rm min}$ and $S<S_{\rm max}$ are required to ensure forward growth devoid of thermal end combinatorial errors. Ultimately, the speed, encoding capacity and error fraction of Eqs.~(\ref{eq:t_ret}-\ref{eq:error}) depend acutely on the number of bonds at critical events, $n_{\rm c}$. In the next two sections, we demonstrate how controlling $n_{\rm c}$ allows to resolve speed and encoding bottlenecks.

\section{Resolving bottlenecks}\label{sec.sol}
To increase robustness in our self-assembly model, we focus on the critical addition events. These typically manifest at assembly boundaries and at the onset of new layers. By contrast, additions that extend a layer tend to create more bonds and thus proceed faster [Eq.~\eqref{eq:rates}]. In this manner, growth tends to proceed layer by layer. naturally favoring layered growth. In the following, we analyze the dependence of the critical number of bonds per addition, $n_{\rm c}$, on the monomer connectivity, $z$.

\begin{figure}[!t] 
    \centering
    \includegraphics[width=\linewidth]{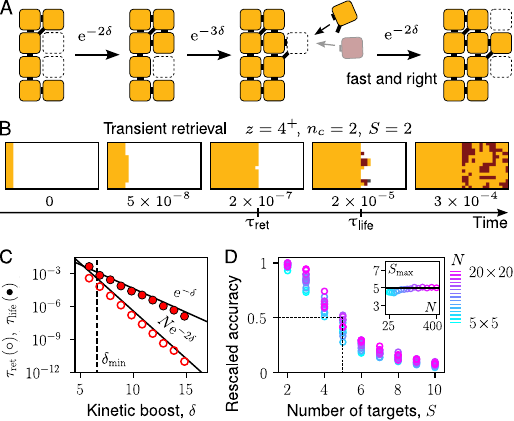}
   \caption{{\it Local increases in connectivity resolve assembly bottlenecks.} 
   {\bf A.} Schematic showing that adding one diagonal bond per layer enables fast and accurate retrieval of $S=2$ targets by ensuring $n_{\rm c}=2$. 
   {\bf B.} Assembly snapshots show faithful retrieval for the same settings as in Fig.~\ref{fig:Z4_S=2}A, except that $z=4^+$. Coloring as in Fig.~\ref{fig:Z4_S=2}A.
   {\bf C.} Retrieval time and lifetime separate at large $\delta$, as predicted. 
   {\bf D.} Accuracy, rescaled between 0 and 1, decreases sigmoidally with $S$. Inset: the sigmoid midpoint $S_{\rm max}$ shows negligible $N$-dependence [Eq.~\eqref{eq:Smax_gen}].}
\label{fig:Z4+}
\end{figure} 

In the case of 1D self-assembly with nearest neighbors interactions, each added monomer forms $n_{\rm c}=1$ bond, which hampers monomer discrimination and limits assembly speed~\cite{Benoist25_SA}. Achieving $n_{\rm c}=2$ requires that each monomer interacts with its next-nearest neighbors, increasing the bulk connectivity from $z=2$ to $4$. By contrast, for quasi-2D assemblies which are the focus of this manuscript, $n_{\rm c}$ can be increased without significantly increasing the connectivity of the whole assembly. Indeed, because the critical events correspond to the creation of new layers, one additional diagonal bond per layer in each target suffices to reach $n_{\rm c}=2$; see Fig.~\ref{fig:Z4+}A. Therefore, among the $N=\ell\times\ell$ assembly components, only $2\ell$ need an increased connectivity $z_i=5$ instead of $z_i=4$, which only marginally increases the average bulk connectivity of structures to $z=4+2/\ell$ (hereafter written as $4^+$). From Eqs.~(\ref{eq:t_ret}-\ref{eq:error}), we expect drastic improvements in retrieval time $\tau_{\rm ret}$, encoding capacity $S_{\rm max}$ and minimum error $e_{\rm min}$, even though the change in connectivity of structures is small.

\begin{figure*}[!t] 
\centering
\includegraphics[width=\linewidth]{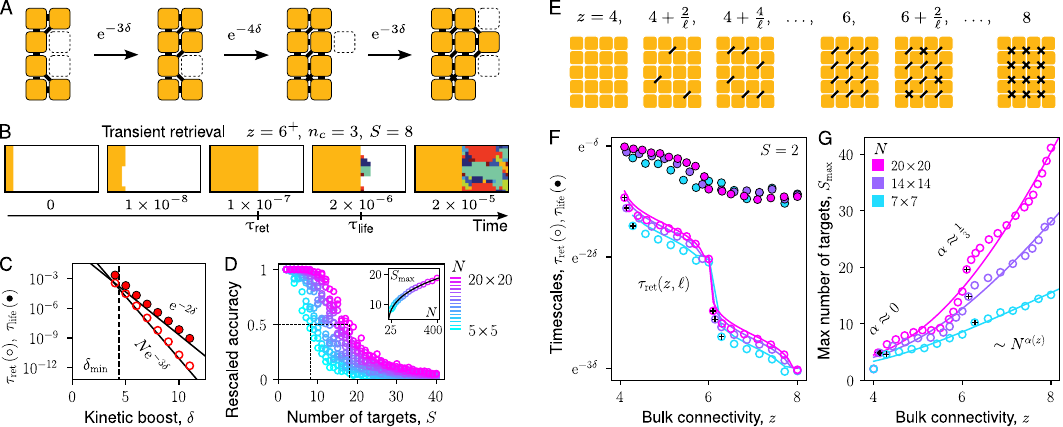}
\caption{{\it Larger increases in connectivity enable combinatorial encoding of target structures.} 
{\bf A.} The connectivity is raised to $z=6^+$, ensuring a minimum number of bonds per addition $n_{\rm c}=3$. 
{\bf B.} Assembly snapshots show faithful retrieval for $S=8$ targets. Here, $\mu=20$ and $\delta=7$.
{\bf C.} Retrieval time and lifetime separate once again at large $\delta$, ensuring kinetic stability.
{\bf D.} For $n_{\rm c}=3$, $S_{\rm max}$ scales as $N^{1/3}$ in agreement with Eq.~\eqref{eq:Smax_gen}.
{\bf E.} To interpolate between the cases $z=4^+$ and $6^+$, we gradually increase the number of diagonal bonds in target structures as shown. (Tile contours and non-diagonal bonds are omitted for clarity.)
{\bf F.} Dependence of timescales on connectivity $z>4$ for $S=2$, $\mu=20$, $\delta=10$ and target side lengths $\ell=7,14,20$. The retrieval time $\tau_{\rm ret}$ follows Eq.~\eqref{eq:t_ret} (lines). Cases with connectivity $z=4^+$ and $6^+$ are indicated as ``+'' markers.
{\bf G.} The encoding capacity increases with target size depending on connectivity. The fits (lines) use the scaling exponent $\alpha=1-4/z$ from Eq.~\eqref{eq:SmaxvZ}.}
\label{fig:Z6+_and_allZ}
\end{figure*}
 
This minor connectivity modification raises the rate of critical additions to $k_{\rm c}=\exp(2\delta)$ and increases their number from $N_{\rm c} = \ell$ to $\ell(\ell - 1)$, so that there are no more speed bottlenecks; see~\cite{SM} Sec.~\ref{sec.SI-Dep}. Moreover, the multiplicity of correct additions at critical events collapses from $\Omega_{\rm C}=\ell$ to $1$ or $2$, while the multiplicity of incorrect additions $\Omega_{\rm I}$ remains unchanged. 
This enables fast and reliable retrieval of the seeded target despite components being reused in $S=2$ targets; see Fig.~\ref{fig:Z4+}B. Separation between retrieval time and lifetime occurs above the reduced discrimination threshold $\delta_{\rm min}=\frac52\ln\ell$ [Fig.~\ref{fig:Z4+}C and Eq.~\eqref{eq:d_min}]. While the lifetime shortens to $\tau_{\rm life}\sim\exp(-\delta)$, retrieval becomes exponentially faster due to the elimination of bottlenecks, consistent with the prediction $\tau_{\rm ret}\approx N\exp(-2\delta)$ [Eq.~\eqref{eq:t_ret}]. Furthermore, Eq.~\eqref{eq:Smax_gen} correctly predicts a non-scaling encoding capacity $S_{\rm max}=\order{1}$ for $n_{\rm c}=2$; see Fig.~\ref{fig:Z4+}D and also Fig.~\ref{fig:error} regarding error fraction.

To conclude, increases in local connectivity at assembly bottlenecks funnels growth toward 
the seeded target. This results in drastic improvements in retrieval speed and accuracy, conferring robustness against moderate component promiscuity.

\section{Combinatorial self-assembly}\label{sec.combi}
In the previous section, we showed that encoding bottlenecks can be overcome by adding an extra bond to $2\ell\ll N$ components. However, the maximal number of stored structures did not scale with $N$. Equation~\eqref{eq:Smax_gen} predicts that the encoding capacity, $S_{\rm max}$, scales with $N$ for $n_{\rm c}\ge3$. In this section, we explore how to achieve this combinatorial scaling in the number of target structures.

To guarantee that $n_{\rm c}=3$, we consider assemblies with a bulk connectivity $z=6$, whereby all components form diagonal bonds with their north-west and south-east neighbors in each target. We then repeat the strategy in the previous section and add an additional bond for one component per layer (in the south-west direction), so that these components and their partners have $z_i=7$. This modification raises the average bulk connectivity to $z=6+2/\ell$ (hereafter denoted $6^+$) and ensures that $n_{\rm c}=3$; see Fig.~\ref{fig:Z6+_and_allZ}A. 

The higher connectivity leads to faithful retrieval for $S\ge2$ targets [Fig.~\ref{fig:Z6+_and_allZ}B] above the reduced discrimination threshold $\delta_{\rm min}\approx\frac53\ln\ell$ [Eq.~\eqref{eq:d_min}]. As shown in Fig.~\ref{fig:Z6+_and_allZ}C, for $S$ near $S_{\rm max}$, the large overlap between targets shortens lifetimes to $\tau_{\rm life}\approx \exp(-2\delta)$, while $\tau_{\rm ret}\approx N\exp(-3\delta)$ [Eq.~\eqref{eq:t_ret}], ensuring kinetic stability of targets. Likewise, the encoding capacity $S_{\rm max}$ follows the predicted scaling $N^{1/3}$ from Eq.~\eqref{eq:Smax_gen} [Fig.~\ref{fig:Z6+_and_allZ}D]. Alternatively, for just $S=2$ targets at $z=6^+$, the error fraction decreases with size as $e_{\rm min}\sim 1/N$ [Eq.~\eqref{eq:error}]; see Fig.~\ref{fig:error}. This validates our predictions and points to robust target designs.

To further assess the effect of connectivity, we interpolate between $z=4^+$ and $6^+$ by gradually adding diagonal bonds in the square lattice; see Fig.~\ref{fig:Z6+_and_allZ}E. Despite this gradual increase in connectivity, the critical number of bonds $n_{\rm c}$ changes discretely, \textit{e.g.} from 2 to 3 as $z$ increases from 6 to $6^+$; see~\cite{SM} Sec.~\ref{sec.SI-Dep}. We thus expect corresponding jumps in the downward trend of retrieval times $\tau_{\rm ret}$ [Eq.~\eqref{eq:t_ret}] and in the upward trend of encoding capacity $S_{\rm max}$. 
Indeed, $\tau_{\rm ret}$ drops sharply with connectivity; see Fig.~\ref{fig:Z6+_and_allZ}F for $S=2$ and Fig.~\ref{fig:tvs&M} for $S$ near $S_{\rm max}$. However, the maximum number of targets $S_{\rm max}$, governed by the scaling exponent $\alpha=1-2/n_{\rm c}$, grows smoothly with $z$; see Fig.~\ref{fig:Z6+_and_allZ}G. Interpolating $n_{\rm c}$ in the expression for $\alpha$ as $z/2$ (\cite{SM} Sec.~\ref{sec.SI-Cond}) yields the combinatorial scaling
\begin{equation}\label{eq:SmaxvZ}
	S_{\rm max}\sim N^{1-4/z} \ \text{ for }\ z\ge4\quad,
\end{equation} 
which captures well the observed increase in encoding capacity with increasing connectivity. This expression shows that for $z=4$, the encoding capacity does not scale with system size, as previously discussed, unlike in previous models in which structures are encoded through binding energies~\cite{Murugan_PNAS15,Sartori20}. Whereas the scalings for $z=6$ and $8$ are $\order{N^{1/3}}$ and $\order{N^{1/2}}$, respectively.

In summary, kinetic encoding enables fast retrieval far from equilibrium. Within a simple quasi-2D geometry, gradually raising the connectivity of structures progressively accelerates retrieval and buffers it against thermal and combinatorial errors.

\section{Discussion}

{\bf Summary.}
In this article we have shown that a large heteromeric structure can rapidly self-assemble from a small seed through kinetic discrimination of its components. This extends our previous work on polymerization~\cite{Benoist25_SA} to the case of quasi-2D growth. Nonetheless, in simple assembly models with fixed connectivity, few specific additions weaken discrimination and decrease assembly speed. These speed and encoding bottlenecks constitute a key difference from 1D growth, where assembly steps have comparable timescales and error rates. We found that adding a single bond per assembly layer alleviates these bottlenecks, yielding faster and more robust assembly. In particular, these localized changes in connectivity greatly reduce errors due to component promiscuity. More broadly, gradual increases in connectivity lead to progressive improvements in assembly speed, accuracy, and encoding capacity.
\newline

{\bf Relationship to previous theoretical work.}
This study builds on recent advances showing that mixtures with many components can encode higher-order structures, both solid~\cite{Murugan_PNAS15,Sartori20} and liquid~\cite{Teixeira24,Jacobs21}, as minima of a high-dimensional free-energy landscape, akin to pattern retrieval in Hopfield neural networks~\cite{Hopfield82,Krotov16,Ramsauer21}. Beyond classical free-energy minimization, achieving rapid far-from-equilibrium growth requires careful design of assembly pathways~\cite{Whitelam15,McMullen22}. 

Within the framework of nonequilibrium self-assembly, selectively accelerating specific binding events emerges as a powerful mechanism to kinetically trap assemblies along pre-encoded target pathways~\cite{Benoist25_SA}. Such kinetic discrimination has proven effective in diverse contexts, including copolymerization~\cite{Banerjee17,Pigolotti16,Bennett79,Tsai06}, minimal DNA self-assembly~\cite{Yin08}, error correction~\cite{Hopfield74,Sartori13,Ravasio24}, random walks~\cite{Lefebvre23} and associative memory~\cite{Benoist25_Hop}. Across these systems, optimal retrieval occurs far from equilibrium, thereby overcoming the intuitive speed-accuracy tradeoff~\cite{Bennett79, Sartori13, Andrieux08}.

Previous self-assembly studies have highlighted the importance of monomer geometry and connectivity~\cite{Gartner24,Lenz17,Manoharan15}. Yet most 2D lattice models, relying on a single monomer type, restrict bulk connectivity to discrete values---triangles bind to three neighbors~\cite{Gartner24}, squares to four~\cite{Nguyen16,Sartori20,Gartner24,Holmes-Cerfon25,Osat23,Bupathy22,Whitelam14}, hexagons to six~\cite{Gartner24,Koehler24}. Our approach goes beyond these geometric constraints: by tuning the connectivity of selected components, we enable a continuous and versatile route toward optimal assembly performance.

Overall, the present work provides a new instance of kinetic encoding, in which kinetic discrimination and tailored connectivity cooperate to robustly guide growth along pre-defined target pathways.
\newline 

{\bf Implications for biological self-assembly.}
As shown in the data analysis of Fig.~\ref{fig:prot_comp}, both kinetic discrimination and variable connectivity are key features of the assembly of large heteromeric protein complexes. Interestingly, several of our results parallel biological assembly principles.

First, the speed bottlenecks that naturally arise in our toy model are similarly observed during \textit{in vitro} ribosome assembly, where they manifest via the accumulation of assembly intermediates~\cite{Seffouh24,Davis17,Mulder10,Sheng23}. Assembly speed and robustness are particularly sensitive to intervention at these bottlenecks. In our toy model, we increased the connectivity of components that participate to these bottlenecks to enhance their kinetic discrimination, leveraging cooperative binding. An alternative strategy is to catalyze these bottlenecks by way of assembly factors (\cite{SM} Sec.~\ref{sec.SI-coop-AF}). This may provide a simple explanation for the increase in the number of assembly factors with assembly size [Fig.~\ref{fig:prot_comp}D]. 
Second, we find that assemblies with low-promiscuity components achieve very low error rates at high connectivity. This feature is especially relevant for ribosomes, whose components are seldom shared with other complexes~\cite{Warner09}, but whose flawless assembly is vital for accurate protein translation~\cite{Alberts15}. 
Third, kinetic encoding, whether arising from cooperative binding events or the action of assembly factors, tends to reduce the number of possible assembly pathways (\cite{SM} Sec.~\ref{sec.SI-path}). This mirrors the ordered assembly observed in many complexes~\cite{Bai19,Sheng23,Lavdovskaia24,Vercellino22,Rousseau18}, where growth proceeds through well-defined intermediates. 

In conclusion, our toy model offers a unifying framework that helps simultaneously explain the high assembly speed, the small number of pathways, and the need for many assembly factors in large heteromeric protein complexes. 
\newline

{\bf Model extensions.} 
We now review some possible extensions of our toy model. The core assumption behind our model is cooperative monomer binding: just as a component with two bonds has twice the binding energy under energetic discrimination, its kinetic barrier shrinks twice under kinetic discrimination. Extensions could explore more general dependencies of binding acceleration on the number of bonds. In the~\cite{SM} Sec.~\ref{sec.SI-Growth}, we in fact discuss the case of saturation in binding acceleration. The model further assumes uniform chemical potential and discrimination strength across component species. We find that small variations of these quantities do not qualitatively alter the main results; see~\cite{SM} Sec.~\ref{sec.SI-Var}. 

Additional extensions include fully two-dimensional growth~\cite{Murugan_PNAS15,Sartori20}, and less constrained off-lattice geometries~\cite{Lenz17}. Other important directions include the incorporation of allosteric interactions~\cite{Metson25} beyond the next-nearest interactions considered here, as well as energetic discrimination mechanisms~\cite{Murugan_PNAS15,Bisker18,Sartori20} alongside the kinetic discrimination studied in this work. More realistic connectivity patterns [Fig.~\ref{fig:prot_comp}E], as well as component diffusion~\cite{Harmon22} and depletion effects~\cite{Murugan_Nat15} also constitute relevant extensions. See extended discussion in~\cite{SM} Sec.~\ref{sec.SI-disc} (and Refs.~\cite{Segel75_effrate,Beck11,Lafontaine21,Lewis00} therein). More broadly, an important direction for future work is to determine how the present results extend to more realistic three-dimensional models of large heteromeric complex self-assembly~\cite{Kurisaki22,Perilla15}.

The catalysis of correct monomer binding events has brought about a new high-speed self-assembly regime. Beyond catalysis, other roles of assembly factors, such as proofreading~\cite{Zhu24,Ravasio24,Sartori13}, could be integrated into future models. More broadly, insights from biological assembly principles may guide the design of synthetic systems. For instance, integrating assembly-factor-like mechanisms into DNA tile assemblies~\cite{Wei12,Evans24,Barish09} could substantially boost both yield and speed.
\\

\acknowledgments{This work was partly supported by a laCaixa Foundation grant (LCF/BQ/PI21/11830032) and core funding from the Gulbenkian Foundation.}


%



\newpage
\widetext
\begin{center}
\textbf{\Large Supporting information for ``Resolving Speed and Encoding \\ \vspace{1.5mm} Bottlenecks in Fast Heteromeric Self-Assembly''}
\end{center}
\setcounter{section}{0}
\setcounter{equation}{0}
\setcounter{figure}{0}
\setcounter{page}{1}
\makeatletter
\renewcommand{\thesection}{S\arabic{section}}
\renewcommand{\theequation}{S\arabic{equation}}
\renewcommand{\thefigure}{S\arabic{figure}}
\renewcommand{\theHfigure}{S\arabic{figure}}

\section{Details of data analysis}\label{sec.SI-Data}
We hereby present details of the data analysis on assembly factors and component connectivity in large heteromeric complexes resulting in the plots of Fig.~\ref{fig:prot_comp}D-E.

{\renewcommand{\arraystretch}{1.}\begin{table}[!h]
	\centering
	\begin{tabular}{|c|c|c|c|}
	\hline
	Complex name & \,Abbreviation in Fig.~\ref{fig:prot_comp}D\, & \,Proteins\, & \,Assembly factors\, \\ \hline
	Spliceosome~\cite{Wahl09} & Splice. & 45 & $125$ \\ \hline
	\ Human ribosome LSU~\cite{Dorner23,Woolford13}\ \ & Rib. LSU & 47 & 91 \\ \hline
	Human ribosome SSU~\cite{Dorner23,Woolford13} & Rib. SSU & 33 & 73 \\ \hline
	Bacterial ribosome~\cite{Dorner23} & B. Rib. & 54 & 50 \\ \hline
	Mitochondrial ribosome~\cite{Lavdovskaia24} & M. Rib. & 82 & 49 \\ \hline
	Mitochondrial complex IV~\cite{Vercellino22} & Comp. IV & 14 & 47 \\ \hline
	Photosystem II (sym.)~\cite{Heinz16} & Ph. II & 20 & 21 \\ \hline
	Mitochondrial complex I~\cite{Vercellino22} & Comp. I & 44 & 20 \\ \hline
	Photosystem I (sym.)~\cite{Yang15} & & 19 & 11 \\ \hline
	RNA polymerase II~\cite{Wild12} & & 12 & 11 \\ \hline
	ATP-synthase (sym.)~\cite{Ruhle15} & & 17 & 10 \\ \hline
	Flagellum (sym.)~\cite{Altegoer15,Armitage20} & & 25 & 8 \\ \hline
	\ Mitochondrial complex III (sym.)~\cite{Vercellino22}\ \ & & 11 & 8 \\ \hline
	Proteasome CP (sym.)~\cite{Rousseau18} & & 14 & 5 \\ \hline
	\ Proteasome RP base (sym.)~\cite{Rousseau18}\ \ & & 9 & 5 \\ \hline
	\end{tabular}
	\caption{{\it Large heteromeric complexes rely on many assembly factors.} We report the approximate number of different proteins and assembly factors regardless of their stoichiometry for a number of large heteromeric complexes. When possible, the assembly factors are distinguished between sub-complexes that assemble separately. This is the case for the ribosome large and small subunits named LSU and SSU. Likewise for the proteasome core protein (CP) and regulatory protein (RP). Local rotational symmetries in the complex are indicated with the abbreviation sym.}
	\label{table_AF}
\end{table}}

\begin{figure}[!h]
    \centering
    \includegraphics[width=.8\linewidth]{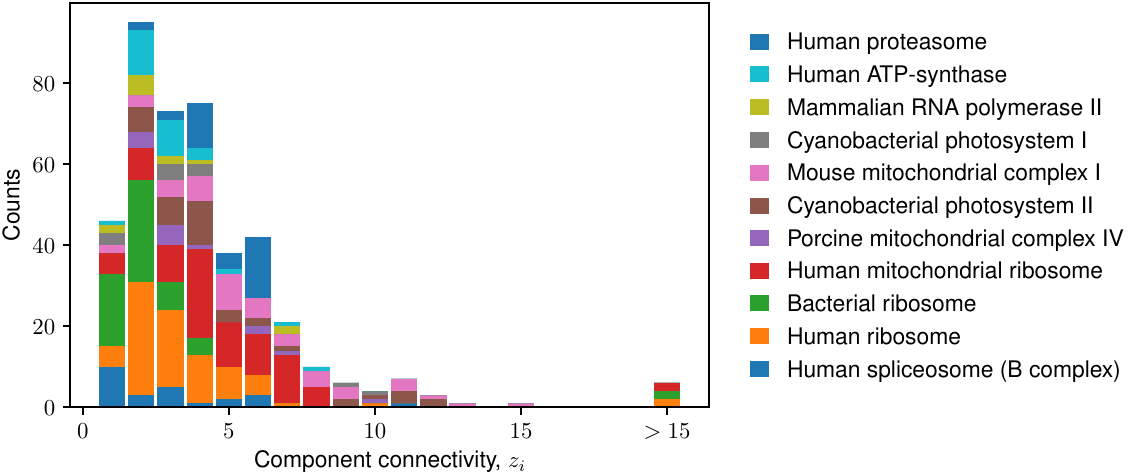}
	\caption{{\it The connectivity of components in large heteromeric complexes is highly variable.} We report the connectivity of components within a number of large heteromeric complexes. Highly-connected components correspond to RNAs. The black bars in Fig.~\ref{fig:prot_comp}D correspond to the human ribosome data (orange bars here). Protein data bank IDs: 6AHD, 4UG0, 7K00, 3J9M, 8UGL, 3WU2, 7B93, 1JB0, 7B7U, 8H9T, 5L4G.}
\label{fig:con_detailed}
\end{figure}

\section{Kinetic discrimination from cooperativity and assembly factors}\label{sec.SI-coop-AF}
Our toy model relies on a minimal description of kinetic discrimination, in which a single parameter $\delta$ captures both kinetic cooperativity and assembly-factor catalysis. This coarse-grained description does not resolve microscopic mechanisms; it isolates their common effect: modulation of kinetic barriers. We now clarify how the framework connects to these two mechanisms.\\

\begin{figure}[!h] 
   \centering
   \includegraphics[width=.75\linewidth]{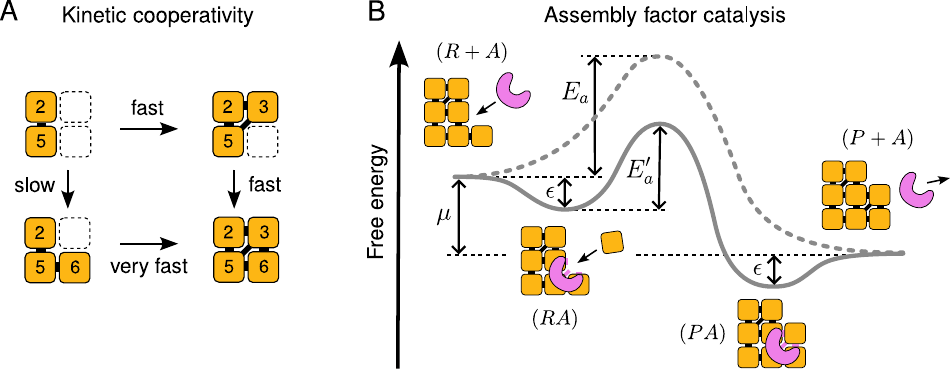}
   \caption{{\it Kinetic discrimination from kinetic cooperativity and catalyzing assembly factors.} {\bf A.} Here, the bindings of components $3$ and $6$ are kinetically cooperative. In our model, binding of component $6$ before $3$ binds creates $r_6 = 1$ bond and has a binding rate $k^+_6=\exp(\delta)$, while binding of component $6$ after $3$ binds creates $r_6 = 2$ bonds and has a faster binding rate $k^+_6=\exp(2\delta)$. Similarly, pre-binding of component 6 accelerates binding of component 3 from $k^+_3=\exp(2\delta)$ to $\exp(3\delta)$. {\bf B.} An assembly factor (pink) catalyzes the binding of a monomer to the growing structure. The free-energy difference between the initial state with 7 monomers and the final state with 8 monomers is the chemical potential $\mu$. Without assembly factor (dashed path), binding involves a large activation energy $E_a$. With assembly factor, stable intermediary states are introduced, which effectively lowers the activation energy to a value $E_a'$ and accelerates binding. (Assembly sketches as in Fig.~\ref{fig:prot_comp}F.)}
\label{fig:catalysis_drawing}
\end{figure}

{\bf Kinetic cooperativity.} 
In heteromeric self-assembly, binding events can be kinetically cooperative: pre-binding of one component can accelerate subsequent binding; see Fig.~\ref{fig:catalysis_drawing}A. During \textit{in vitro} ribosome assembly, binding of primary proteins to ribosomal RNA accelerates binding of secondary proteins by up to two orders of magnitude~\citep{Bunner10}, likely through increased interfaces or conformational changes~\citep{Williamson08}. We model this effect by assigning each binding event a number of correct bonds $r_i$, and defining the rate $k^+_i=\exp(r_i\delta)$ [Eq.~\eqref{eq:rates}], so that bindings forming more correct contacts are exponentially accelerated. The parameter $\delta$ thus encodes kinetic discrimination between competing binding events.
\\

{\bf Assembly factor catalysis.} 
Enzymatic catalysis corresponds to a reduction in activation energy~\citep{Alberts15}. Since assembly factors are not part of the final complex, they do not alter binding energies; instead, they lower the activation barrier (in units of $k_{\rm B}T$) from $E_a$ to $E_a'=E_a-\delta$; see Fig.~\ref{fig:catalysis_drawing}B. 
Without assembly factor, the forward and backward rates are simply written
\begin{equation}\label{k+-_Ea}
	k_+ \propto\exp(-E_a) \qq{and} k_-\propto\exp(-E_a-\mu).
\end{equation}
With an assembly factor, we consider the reaction network
\begin{equation}
	\ce{ $R+A$ <=>[$k_1$][$k_{-1}$] $RA$ <=>[$k_2$][$k_{-2}$] $PA$ <=>[$k_3$][$k_{-3}$] $P+A$ },
\end{equation}
connecting the reactant $R$ (assembly with 7 monomers) to the product $P$ (assembly with 8 monomers) via an assembly factor $A$ and two intermediates. We express the rates up to a time contant from the energy barriers in Fig.~\ref{fig:catalysis_drawing}B as 
\begin{equation}
\begin{matrix}
k_1=1 \qquad\  \\
k_{-1}=\exp(-\epsilon)
\end{matrix}\ ,\qquad
\begin{matrix}
k_2=\exp(-E_a') \quad\, \\
k_{-2}=\exp(-E_a'-\mu)
\end{matrix}\ ,\qquad
\begin{matrix}
k_3=\exp(-\epsilon) \\
k_{-3}=1 \qquad\quad\ \
\end{matrix}.
\end{equation}
This is such that, \textit{e.g.}, $\frac{\rm d}{{\rm d}t}[R]=k_{-1}[RA]-k_1[R][A]$, where all concentrations are dimensionless. Applying steady-state elimination of intermediates~\citep{Segel75_effrate} and denoting by $[A]_t$ the total concentration of assembly factors (present either in form $A,RA$ or $PA$), the effective rates between $R$ and $P$ read
\begin{equation}
	k'_+ =\frac{k_1k_2k_3[A]_t}{X}=\frac{\exp(-E_a'-\epsilon)[A]_t}{X} \qq{and} k'_- =\frac{k_{-1}k_{-2}k_{-3}[A]_t}{X}=\frac{\exp(-E_a'-\mu-\epsilon)[A]_t}{X} \quad,
\end{equation}
with the common denominator
\begin{equation}
\begin{aligned}
	X &= k_{-1}k_{-2}+k_{-1}k_3+k_2k_3 + k_1(k_{-2}+k_2+k_3)[R] + k_{-3}(k_{-1}+k_{-2}+k_2)[P] \\
	  &= \Big(\exp(-E_a'-\mu)+\exp(-E_a')+\exp(-\epsilon)\Big)\Big([R]+[P]+\exp(-\epsilon)\Big)\quad.
	\end{aligned}
\end{equation}
For concentrations of $R$ and $P$ of order unity and at low temperature, one has $[R]+[P]\gg\exp(-\epsilon)$, while in the limit of large activation energy, $E_a'\gg \epsilon$. The denominator therefore reduces to $X\approx \exp(-\epsilon)([R]+[P])$. The factor $\exp(-\epsilon)$ then cancels that in the numerators yielding
\begin{equation}
	k'_+ \approx \frac{[A]_t}{[R]+[P]}\exp(-E_a') \qq{and} k'_-\approx\frac{[A]_t}{[R]+[P]}\exp(-E_a'-\mu),
\end{equation}
which recovers Eq.~\eqref{k+-_Ea} with a reduced activation barrier. The prefactor indicates that the rates also increase with the concentration of assembly factors.

Thus, assembly factors accelerate both forward and backward rates by a factor $\exp(\delta)$. Since assembly factors bind at specific locations and times during complex growth (Fig.~\ref{fig:prot_comp}B–C), their catalytic effect depends on the identity of the interacting components. As a result, assembly factor action effectively selects specific binding events, acting analogously to kinetic cooperativity by kinetically discriminating between binding partners.
\\

{\bf Combined framework.}
Within this framework, $\delta$ refers just as well to kinetic cooperativity as to assembly factor catalysis. In this sense, assembly factors effectively enable specific components to access otherwise less favorable binding configurations. As shown in the main text, increasing the connectivity $z_i$ of only a subset of components, \textit{i.e.} those involved in rate-limiting steps, substantially accelerates assembly. Alternatively, assembly factors may have evolved to catalyze the critical addition steps, consistent with experimental observation.

\section{Methods for Monte-Carlo simulations}\label{sec.SI-Methods}
In our simulations, the boundary-growth process is modeled as follows. Removal of bound monomers is permitted only at the rightmost side of the assembly. This corresponds to locations $\mathbf x=(x,y_-(x))$ with vertical coordinate $x=1,\dots,\ell$ and horizontal coordinate $y=y_-(x)$, the maximal value occupied by a monomer. For instance, in Fig.~\ref{fig:sketch}B, $y_-(1)=2$ so $\mathbf x=(1,2)$ is a removal location. Addition locations are defined analogously as $\mathbf x=(x,y_+(x))$ with $y=y_+(x)=y_-(x)+1$, corresponding to the minimum unoccupied value. To maintain a relatively smooth boundary, we constrain overhangs to a maximum of three. This construction yields two sets of removal and addition locations, $\mathcal X_\pm$, each containing at most $\ell$ locations. 

\begin{figure}[!h]
    \includegraphics[width=.5\linewidth]{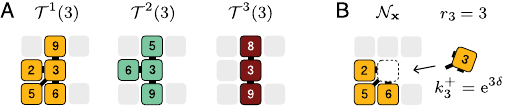}
   \caption{{\it Neighborhood in targets and at binding locations.} 	
   {\bf A.} For each target in Fig.~\ref{fig:sketch}A, component 3 has several encoded neighbors, summarized in the target neighborhoods $\mathcal T^\alpha(3)$, $\alpha=1,2,3$. For example, in the yellow target, $\mathcal T^1(3)=(0,9,0,2,5,6,0)$, taking into account periodic vertical boundaries. Here, 0 denotes empty sites. 
   {\bf B.} Its addition at location $\mathbf x$ depends on the neighborhood $\mathcal N_{\bf x}=(0,0,0,2,5,6,0)$. Three of these neighbors match the encoded neighbors of component 3, giving $r_3=3$ [Eq.~\eqref{eq:r_ix}].} 
\label{fig:neighborhoods}
\end{figure}

Then, we calculate the binding/unbinding rates of Eq.~\eqref{eq:rates} as follows. Each target structure labeled $\alpha=1,\dots,S$ is encoded locally through specific target neighborhoods for each component $i=1,\dots,N$ written $\mathcal T^\alpha(i)=(i_{\rm ne}^\alpha,\dots,i_{\rm se}^\alpha)$; see Fig.~\ref{fig:neighborhoods} for examples. For a given component $i$ at location $\bf x$, the number of bonds $r_i(\mathcal{N}_{\bf x})$ equals the number of neighbors in $\mathcal N_{\bf x}=(i_{\rm ne},\dots,i_{\rm se})$ that also appear in at least one target neighborhood $\mathcal T^\alpha(i)$. For example, neighbor $i_{\rm ne}$ is compared against all encoded neighbors $i_{\rm ne}^1,\dots,i_{\rm ne}^S$. This is formalized as
\begin{equation}\label{eq:r_ix}
    r_i(\mathcal{N}_{\bf x})=\sum_{\theta={\rm ne}}^{\rm se}\Theta\bigg(\sum_{\alpha=1}^Sf(i_\theta,i_\theta^\alpha)\bigg)\quad, \qq{where}
    f(i,j)= \begin{cases}
	1,&\text{if }\ i=j\neq0\\
	0,&\text{otherwise}\end{cases}
\end{equation}
and $\Theta$ is the Heaviside function. While $r_i(\mathcal{N}_{\bf x})$ may in principle reach 7, the connectivity $z_i$ of components in targets limits the number of bonds. For instance, in Fig.~\ref{fig:neighborhoods}, component 3 binds to neighbors $i_{\rm w}=i_{\rm w}^1=2$, $i_{\rm sw}=i_{\rm sw}^1=5$ and $i_{\rm s}=i_{\rm s}^1=6$, yielding $r_3=3$ bonds (black rectangles), and thus an addition rate $k_3^+=\exp(3\delta)$, according to Eq.~\eqref{eq:rates}.

To simulate stochastic growth, we use the Gillespie algorithm. Starting from a nucleation seed, each iteration consists of four tasks: (i) We list all possible additions, $\mu=(i,\bf x)$ with $i=1,\dots,N$ and $\mathbf x\in\mathcal X_+$, and all possible removals $\mu=(\mathbf x)$ with $\mathbf x\in\mathcal X_-$. (ii) We compute their rates $k_\mu$ using Eq.~\eqref{eq:rates}, which sum to $k_{\rm out}=\sum_\mu k_\mu$. (iii) The waiting time is drawn from an exponential distribution with mean $1/k_{\rm out}$. (iv) We select an event $\nu$ with probability $k_\nu/k_{\rm out}$ and update the system accordingly.

\newpage

\section{Principal growth regimes in parameter sweeps}\label{sec.SI-Growth}
We next discuss the growth regimes arising for different values of $\mu$ and $\delta$. We begin with the case of a single square target with tiles of connectivity $z=4$, as in Fig.~\ref{fig:Z4}. We discuss tile fluctuations and saturation in $r_i(\mathcal{N}_{\bf x})$, and then turn to growth regimes for multiple targets and higher connectivity.
\newline

\begin{figure}[!b]
    \centering
    \includegraphics[width=\linewidth]{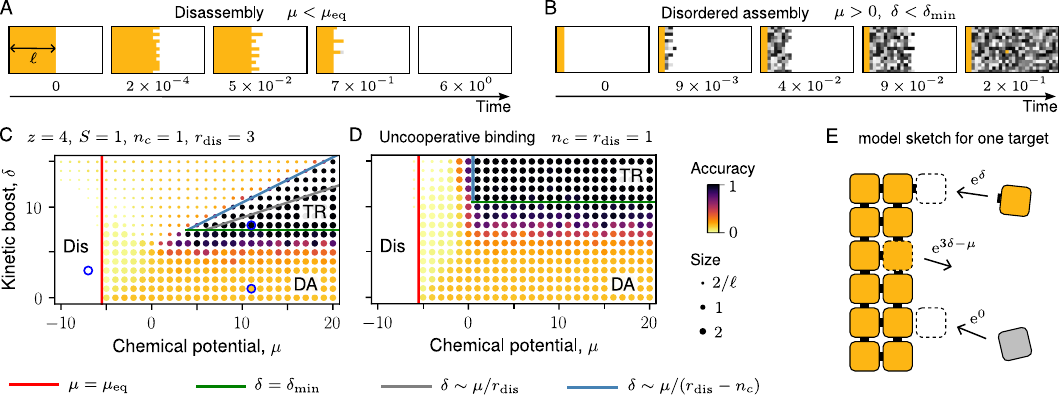}
	\caption{{\it Growth regimes for a single target.} 
	{\bf A.} Snapshots of disassembly for $S=1$ target with $\ell=14$ and connectivity $z=4$. The maximum-overhang rule constrains the shape of the boundary. 
	{\bf B.} Snapshots of disordered growth. 
	{\bf C.} Simulation outcomes after $10^6$ events, starting from a two-layer seed. The accuracy is shown as marker color and the assembly size as marker size. The red line denotes the equilibrium threshold [Eq.~\eqref{eq:mu_red}], the green line denotes the discrimination threshold [Eq.~\eqref{eq:d_min}], and the gray and blue lines satisfy Eq.~\eqref{eq:delta_g_b}. Blue circles refer to the assembly snapshots shown in panels {\bf A}, {\bf B} and Fig.~\ref{fig:Z4}A. 
	{\bf D.} Same as {\bf C}, but for uncooperative binding [Eq.~\eqref{eq:rates1d}]. This results in $n_{\rm c}=1$, $N_{\rm c}=N=\ell^2$, $\Omega_{\rm C}=\ell$ and thus $\delta_{\rm min}=4\ln\ell$. Here, more assemblies have reached size $2$ within $10^6$ events. 
	{\bf E.} Key events at layer onset: correct addition with $n_{\rm c}=1$ bond, incorrect addition with $0$ bonds, and removal breaking $r_{\rm dis}=3$ bonds.}
\label{fig:Z4-SI}
\end{figure}

{\bf Growth regimes for $S=1$ and $z=4$.}
Beyond the transient retrieval (TR) regime for $\mu>0$ and $\delta>\delta_{\rm min}$ [Fig.~\ref{fig:Z4}A], we identify two classical growth regimes in other parts of parameter space. For chemical potentials below the equilibrium threshold~\cite{Bennett79,Sartori13,Andrieux08} 
\begin{equation}\label{eq:mu_red}
	\mu_{\rm eq}=-\ln N \quad, 
\end{equation}
even the full target disassembles (Dis); see Fig.~\ref{fig:Z4-SI}A. For large positive $\mu$, assemblies fill the lattice; but when $\delta<\delta_{\rm min}$ [Eq.~\eqref{eq:d_min}], they become disordered (DA); see Fig.~\ref{fig:Z4-SI}B. These regimes closely mirror those found in the previous polymerization study~\cite{Benoist25_SA} and are recapitulated in the diagrams of Fig.~\ref{fig:Z4-SI}C-D. Importantly, transient retrieval occurs only for $\mu>0$ far from equilibrium [Eq.~\eqref{eq:mu_red}], for large positive values of $\mu$, whereas retrieval of assemblies stabilized by energetic couplings occurs for $\mu<0$ near equilibrium~\cite{Sartori20}. Thus, the conditions for accurate growth under kinetic versus energetic encoding are fundamentally distinct, consistent with previous copolymerization results~\cite{Sartori13}.

Interestingly, Fig.~\ref{fig:Z4-SI}C reveals a large fraction of the nominal retrieval region, $\mu>0$ and $\delta>\delta_{\rm min}$, where assemblies stall. Above the blue line, growth halts near the seed size ($2/\ell$), while between the gray and blue lines, the target is recovered without additional incorrect layers. Both effects stem from the imbalance between layer-closing and layer-nucleating events: closures involve more bonds and are accelerated, causing persistent fluctuations in the presence of high $\delta$ and low $\mu$. To rationalize these regimes, we compare binding and unbinding rates [Eq.~\eqref{eq:rates}]. At the onset of each new layer [Fig.~\ref{fig:Z4-SI}E], a correct addition occurs if $k_{\rm c}=\exp(\delta)\gtrsim \exp(3\delta-\mu)$, which facilitates subsequent additions with rate $\exp(2\delta)$. If instead $k_{\rm c}\lesssim \exp(3\delta-\mu)$, repeated additions/removals dominate, creating long-lived fluctuations. This yields the approximate boundary $\delta_{\rm blue}\sim \mu/2$ (blue line in Fig.~\ref{fig:Z4-SI}C). For $\delta$ between $\delta_{\rm min}$ and $\delta_{\rm blue}$, once the target is completed, additional chimeric growth is suppressed provided $\exp(0)=1\gg \exp(3\delta-\mu)$, giving the gray threshold $\delta_{\rm gray}\sim\mu/3$. Since these fluctuation-driven events are slow, both gray and blue boundaries shift upward for longer simulations.
\newline

{\bf Bond saturation.}
To test robustness, we imposed uncooperative binding by saturating $r_i(\mathcal{N}_{\bf x})$ to 1, so each monomer binds to at most one neighbor. The resulting rates are
\begin{equation}\label{eq:rates1d}
	\left\{\begin{aligned}k_i^+(\mathcal N_{\bf x})&=\exp(\delta)\  \text{ and }\ k_i^-(\mathcal N_{\bf x})=\exp(\delta-\mu),\ \text{ for }\ r_i(\mathcal{N}_{\bf x})>0 \quad,\\ 
	k_i^+(\mathcal N_{\bf x})&=1 \quad\quad\ \ \text{ and }\ k_i^-(\mathcal N_{\bf x})=\exp(-\mu),\quad\: \text{ for }\ r_i(\mathcal{N}_{\bf x})=0 \quad,\end{aligned}\right.
\end{equation}
which reduces the range of timescales. Target retrieval remains possible [Fig.~\ref{fig:Z4-SI}D], though only in the $S=1$ case since $n_{\rm c}=1$. Here, the blue line is vertical, and the gray line (slope 1) emerges only at larger $\delta$. In general, for critical events with $n^t=n_{\rm c}$ and removals breaking $r_{\rm dis}$ bonds, the region where the lattice is filled and the retrieval region have respective upper bounds 
\begin{equation}\label{eq:delta_g_b}
	\delta_{\rm gray}\sim \mu/r_{\rm dis} \qq{and} \delta_{\rm blue}\sim \mu/(r_{\rm dis}-n_{\rm c}) \quad. 
\end{equation}
Saturating $r_{\rm dis}$ to a value near $n_{\rm c}$ enlarges the region between gray and blue lines. This can be an efficient strategy for rapid but accurate assembly with slow addition of extra target chunks.   
\newline

\begin{figure}[!h]
    \centering
    \includegraphics[width=\linewidth]{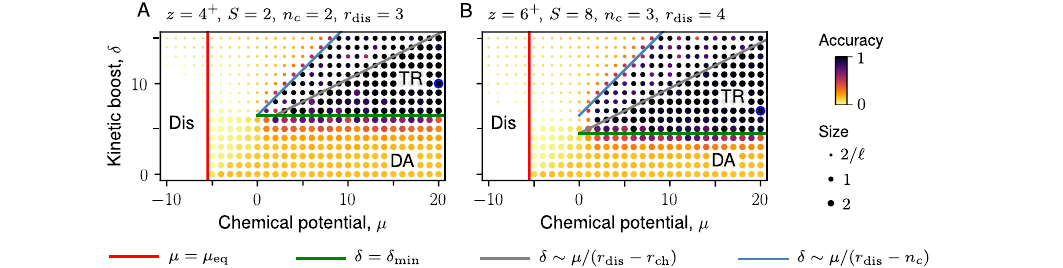}
   \caption{{\it Growth regimes for higher connectivity below encoding capacity.} {\bf A.} Simulation outcomes after $10^6$ events, starting from a two-layer seed and for $z=4+2/\ell$ and $S=2$. The gray line with slope $\frac12$ yields $r_{\rm ch}=1$, consistent with $\tau_{\rm life}\approx\exp(-\delta)$ in Fig.~\ref{fig:Z4+}C. The blue circle refers to the snapshots in Fig.~\ref{fig:Z4+}B. {\bf B.} Additional simulation outcomes for $z=6+2/\ell$ and $S=8$. The gray line with slope $\frac12$ yields $r_{\rm ch}=2$, consistent with $\tau_{\rm life}\approx\exp(-2\delta)$ in Fig.~\ref{fig:Z6+_and_allZ}C. The blue circle refers to the snapshots in Fig.~\ref{fig:Z6+_and_allZ}B.}
\label{fig:diags_varZ}
\end{figure}

\begin{figure}[!h]
    \centering
    \includegraphics[width=\linewidth]{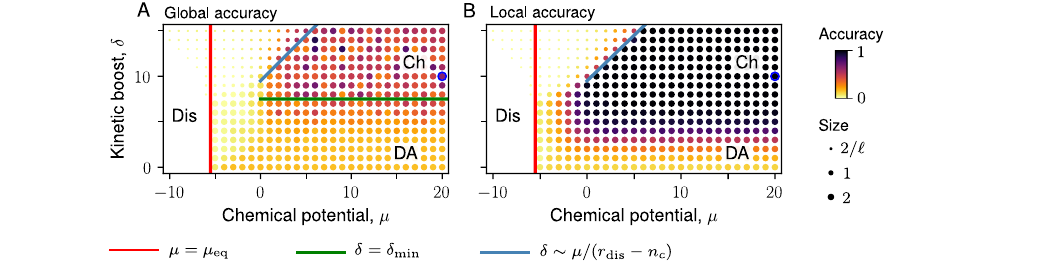}
   \caption{{\it Chimeric regime above encoding capacity.} Simulation outcomes after $10^6$ events, starting from a two-layer seed. Here, $z=4$, $S=S_{\rm max}=2$, and $r_{\rm dis}=2$. 
   	{\bf A.} Global accuracy with the seeded target saturates near $\frac12$.
	{\bf B.} Local accuracy reaches 1, consistent with the snapshots in Fig.~\ref{fig:Z4+}C (blue circle).}
\label{fig:diags_chim}
\end{figure}

{\bf $S>1$ and $z\ge4$.}
With multiple targets but below encoding capacity, $S<S_{\rm max}$, the same growth regimes appear; see Fig.~\ref{fig:diags_varZ}. The boundaries follow the predictions, except for the gray line. Here, overlapping targets accelerate the addition of extra monomers, with rate $\exp(r_{\rm ch}\delta)\sim 1/\tau_{\rm life}$, where $r_{\rm ch}$ grows with component promiscuity, \emph{i.e.} $S$. This shifts the upper bound for target completion to
\begin{equation}\label{eq:d_gray_reuse}
	\delta_{\rm gray}\sim \mu/(r_{\rm dis}-r_{\rm ch}) \quad. 
\end{equation}	
At and above encoding capacity, $S\ge S_{\rm max}$, the retrieval regime disappears, replaced by a chimeric regime (Ch) characterized by high local but low global accuracy; see definitions in main text Sec.~\ref{sec.model} and Fig.~\ref{fig:diags_chim}. 

\newpage

\section{Dependencies with connectivity}\label{sec.SI-Dep} 
We now examine how $n_{\rm c}$, $N_{\rm c}$ and $\Omega_{\rm C}$ depend on the number of extra bonds in the square-lattice case [Fig.~\ref{fig:Z6+_and_allZ}E]. This yields a detailed expression for $\tau_{\rm ret}$ following Eq.~\eqref{eq:t_ret}. 

{\renewcommand{\arraystretch}{1.5}\begin{table*}[!h]
	\centering
	\begin{tabular}{|c|c|c|c|c|c|c|c|c|}
	\hline
	\ \ Number of extra bonds per layer, $s\ \ $ &$\ 0\ $&$\ 1\ $&$\ 2\ $&$\ \dots\ $&$\ \ell-1\ $&$\ \ell\ $&$\ \ell+1\ $&$\ \dots\ $\\ \hline
	Bulk connectivity, $z$ &$\ 4\ $&$\ 4+\frac2\ell\ $&$\ 4+\frac4\ell\ $&$\ \dots\ $&$\ 6-\frac2\ell\ $&$\ 6\ $&$\ 6+\frac2\ell\ $&$\ \dots\ $\\ \hline
	Critical number of bonds, $\ n_{\rm c}\ $&$\ 1\ $&$\ 2\ $&$\ 2\ $&$\ \dots\ $&$\ 2\ $&$\ 2\ $&$\ 3\ $&$\ \dots\ $\\ \hline
	Approximation for $n_{\rm c}$, $\ 2+\frac{s-1}\ell\ $&$\ 2-\frac1\ell\ $&$\ 2\ $&$\ 2+\frac1\ell\ $&$\ \dots\ $&$\ 3-\frac2\ell\ $&$\ 3-\frac1\ell\ $&$\ 3\ $&$\ \dots\ $\\ \hline
	Number of critical events (CE), $N_{\rm c}\ $&$\ \ell\ $&$\ \ell(\ell-1)\ $&$\ \ell(\ell-2)\ $&$\ \dots\ $&$\ \ell\ $&$\ \ell\ $&$\ \ell(\ell-1)\ $&$\ \dots\ $\\ \hline
	Multiplicity of correct additions at CE, $\Omega_{\rm C}\ $&$\ \ell\ $&$\ 1$-$2\ $&$\ 2$-$4\ $&$\ \dots\ $&$\ \ell-1\ $&$\ \ell\ $&$\ 1$-$2\ $&$\ \dots\ $\\ \hline
	Discrimination threshold, $\delta_{\rm min}\ $&$\ 3\ln\ell\ $&$\ \frac52\ln\ell\ $&$\ \frac52\ln\ell\ $&$\ \dots\ $&$\ \frac32\ln\ell\ $&$\ \frac32\ln\ell\ $&$\ \frac53\ln\ell\ $&$\ \dots\ $\\ \hline
	Target retrieval time, $\tau_{\rm ret}\ $&$\ {\rm e}^{-\delta}\ $&$\ \ell(\ell-1){\rm e}^{-2\delta}\ $&$\ \frac{\ell(\ell-2)}{2}{\rm e}^{-2\delta}\ $&$\ \dots\ $&$\ \frac{\ell}{\ell-1}{\rm e}^{-2\delta}\ $&$\ {\rm e}^{-2\delta}\ $&$\ \ell(\ell-1){\rm e}^{-3\delta}\ $&$\ \dots\ $\\ \hline
	\end{tabular}
	\caption{{\it Approximate expressions for gradually increasing connectivity.} Extra bonds are added following the sketches in Fig.~\ref{fig:Z6+_and_allZ}E, which progressively increases the bulk connectivity $z$ following Eq.~\eqref{eq:z_def}. At the same time, the critical number of bonds $n_{\rm c}$ jumps between integer values according to Eq.~\eqref{eq:nmin_def}. Due to the smooth increase of exponent $\alpha=1-2/n_{\rm c}$, we approximate $n_{\rm c}$ in $\alpha$ as a smooth function of $s$; see Eq.~\eqref{eq:nmin_app}. Jumps also occur in $N_{\rm c}$ and $\Omega_{\rm C}$ [Eqs.~(\ref{eq:Nc}-\ref{eq:Omegac})], and in $\delta_{\rm min}\approx\ln(\ell^3N_{\rm c}/\Omega_{\rm C})/n_{\rm c}$ and $\tau_{\rm ret}\approx N_{\rm c}/\Omega_{\rm C}\exp(-n_{\rm c}\delta)$ [Eqs.~(\ref{eq:d_min}-\ref{eq:t_ret})].}  
	\label{table}
\end{table*}}

We consider targets made of $\ell$ layers of $\ell$ tiles, for a total size $N=\ell^2$. Adding $s$ extra diagonal bonds per layer to the bulk of an assembly with baseline connectivity $z=4$ yields an average bulk connectivity 
\begin{equation}\label{eq:z_def}
	z=4+2s/\ell \quad;
\end{equation}
as shown in Table~\ref{table}. While $s=0$ relates to nearest-neighbor couplings on the square lattice ($z=4$), $s=\ell$ correspond to nearest-neighbor couplings on the hexagonal lattice ($z=6$), and $s=2\ell$ corresponds to next-nearest-neighbor couplings on the square lattice ($z=8$).

\begin{figure}[!h] 
	\centering
    \includegraphics[width=.425\linewidth]{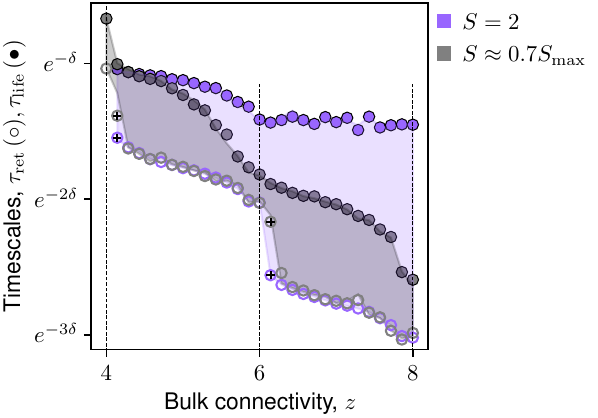}
   \caption{{\it Component reuse reduces timescale separation.} From $S=2$ to $S_{\rm max}$, retrieval times remain mostly independent of $S$, consistent with Eq.~\eqref{eq:t_ret}: Adding more targets does not so much increase $\tau_{\rm ret}$ as prevent retrieval altogether. Whereas $\tau_{\rm life}$ drastically decreases due to increased similarity between targets. Here, $\ell=14$, $\mu=20$ and $\delta=10$.}
\label{fig:tvs&M}
\end{figure}

We now provide expressions for $n_{\rm c}, N_{\rm c}$ and $\Omega_{\rm C}$ which fit the discrete evolution of these quantities with $s$, shown in Table~\ref{table}. Therein, we account for the large jumps via the ceiling function $\lceil .\rceil$. In this setting, the lowest number of bonds per addition increases discretely as 
\begin{equation}\label{eq:nmin_def}
	n_{\rm c}=1+\lceil s/\ell\rceil \quad.
\end{equation} 
The number of critical events with $n^t=n_{\rm c}$ depends on $s$ as 
\begin{equation}\label{eq:Nc}
	N_{\rm c}\approx\max(\lceil s/\ell\rceil\ell-s,1)\ell \quad,
\end{equation}
and at each critical event, correct additions have an approximate multiplicity 
\begin{equation}\label{eq:Omegac}
	\Omega_{\rm C}\approx\ell+s-\lceil s/\ell\rceil\ell \quad.
\end{equation}
Note, however, that the multiplicity $\Omega_{\rm C}$ can vary across different critical events; see Figs.~\ref{fig:Z4+}A,~\ref{fig:Z6+_and_allZ}A and Table~\ref{table}. Inserting these expressions into Eq.~\eqref{eq:t_ret} for the retrieval time $\tau_{\rm ret}$ produces the analytical curves shown in Fig.~\ref{fig:Z6+_and_allZ}F. The dependence of target lifetimes $\tau_{\rm life}$ is analyzed in Fig.~\ref{fig:tvs&M}. 

\section{Conditions for faithful retrieval}\label{sec.SI-Cond} 
We now provide details on the analytical calculations of Sec.~\ref{sec.theory} by examining the error rate at critical events, where errors are most likely to occur. We first derive the conditions for retrieving a single two-dimensional structure, following a previous polymerization study~\cite{Benoist25_SA}. This leads to Eq.~\eqref{eq:d_min} for $\delta_{\rm min}$. We then turn to the case of component reuse across $S$ different structures, obtaining Eqs.~\eqref{eq:Smax_gen} and~\eqref{eq:SmaxvZ} for $S_{\rm max}$ and Eq.~\eqref{eq:error} for $e_{\rm min}$. 
\newline

{\bf Error rate at critical events.}
In contrast to one-dimensional polymerization, higher-dimensional assemblies grow via addition events where monomers can bind to a variable number of neighbors. We focus on the quasi-irreversible growth regime, \emph{i.e.} large $\mu$, where additions dominate over removals, according to Eq.~\eqref{eq:rates}. In this regime, each boundary configuration admits roughly $N\ell=\ell^3$ possible additions: placing any component $i=1,\dots,N$ at any boundary location of vertical coordinate $x=1,\dots,\ell$, except for overhangs; see Sec.~\ref{sec.SI-Methods}. Since there are $N-1$ incorrect monomers at each location, successful retrieval of the target requires strong discrimination, \textit{i.e.} large $\delta$. 

Starting from a seed of size $N_{\rm seed}$, we consider the sequential addition of correct monomers up to the target size $N$, indexing the $N_{\rm tot}=N-N_{\rm seed}$ addition events by $t=1,\dots,N_{\rm tot}$. At each event, the chosen monomer $i^t$ tends to bind to the boundary location $\mathbf x^t$ maximizing the addition rate, as in Eq.~\eqref{eq:it,xt}. We denote this maximum by $k^t=\exp(n^t\delta)$, with $n^t$ the effective number of bonds in the irreversible, high-discrimination regime at time $t$. Events with the smallest number of bonds, $n_{\rm c}=\min_t n^t$, are thus critical [Eq.~\eqref{eq:kc,nc}]. The number of such critical events $N_{\rm c}$ depends on the connectivity of components $z$; see Sec.~\ref{sec.SI-Dep}.  

At each critical event with time index $t'=1,\dots,N_{\rm c}$ and $n^{t'}=n_{\rm c}$, potential correct/incorrect additions correspond to sums of rates $K_{\rm C}(t')$ and $K_{\rm I}(t')$. In the irreversible high-discrimination regime, the probability to retrieve the seeded target is the probability of avoiding errors across all $N_{\rm c}$ critical events:
\begin{equation}\label{eq:Pret_gen}
    P_{\rm ret}\approx \prod_{t'=1}^{N_{\rm c}}\left(1-\frac{K_{\rm I}(t')}{K_{\rm C}(t') + K_{\rm I}(t')}\right) \approx (1-p_{\rm err})^{N_{\rm c}} \quad.
\end{equation}
Here, $p_{\rm err}$ represents the average error rate at critical events, defined as the probability that there is ambiguity about which component to add:
\begin{equation}\label{eq:p_err_gen}
	p_{\rm err}\approx\frac1{N_{\rm c}}\sum_{t'=1}^{N_{\rm c}}\frac{K_{\rm I}(t')}{K_{\rm C}(t')+K_{\rm I}(t')} \quad.
\end{equation} 
Thus, if correct additions dominate ($K_{\rm C}\gg K_{\rm I}$), then $p_{\rm err}\approx0$ and the assembly is nearly error-free. More precisely, a high retrieval probability $P_{\rm ret}$ requires the low-error condition $N_{\rm c} p_{\rm err}\ll1$. 

In what follows, we discard the dependence of $K_{\rm C}$ and $K_{\rm I}$ on the time $t'$ and thus consider their average values over all critical events. We moreover consider the large-$\delta$ limit, and assume that the assembly follows the seeded target, thereby allowing us to approximate $K_{\rm C}$ and $K_{\rm I}$ by their most accelerated terms. This is such that, \textit{e.g.} in Fig.~\ref{fig:sketch}C at time $t=2$, the correct binding of component 5 to components 4 and 7 with rate $\exp(2\delta)$ wins over the other correct binding of 8 to 7 with rate $\exp(\delta)$. We first discuss conditions for the retrieval of a single structure, and then turn to the case of multiple structures.
\newline

{\bf Retrieval of a single target structure.}
In the case of $S=1$ fully heteromeric target, there is only one correct monomer per location. Thus, the lower bound on $\mu$ shifts from its equilibrium value $-\ln N$ [Eq.~\eqref{eq:mu_red}] to $-\ln 1=0$. Next, we estimate the correct rate as
\begin{equation}\label{eq:K_C}
	K_{\rm C}\approx\Omega_{\rm C}\exp(n_{\rm c}\delta) \quad.
\end{equation} 
This corresponds to additions creating $n_{\rm c}$ bonds, and with multiplicity $\Omega_{\rm C}\le \ell$ depending on the connectivity of components $z$; see Sec.~\ref{sec.SI-Dep}.
Since each component is used only once, no incorrect addition benefits from acceleration, giving $K_{\rm I}=\Omega_{\rm I}$. Contrary to $\Omega_{\rm C}$, the multiplicity of incorrect additions is independent of $z$ and reads $\Omega_{\rm I}\approx \ell(N-1)\approx\ell^3$. The resulting rate of thermal errors is
\begin{equation}\label{eq:p_err_delta}
	p_{\rm err}\approx\frac{\Omega_{\rm I}}{\Omega_{\rm C}\exp(n_{\rm c}\delta)+\Omega_{\rm I}} \quad.
\end{equation} 
Imposing $N_{\rm c}p_{\rm err}\ll1$ yields the discrimination threshold $\delta_{\rm min}$ required for correct assembly [Eq.~\eqref{eq:d_min}].
\newline

\begin{figure}[!h]
    \includegraphics[width=.5\linewidth]{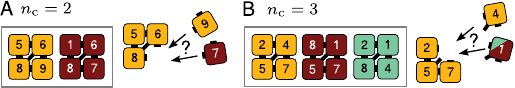}
   \caption{{\it Combinatorial errors.} 
   {\bf A.} Binding event with impossible discrimination. Failure to retrieve the seeded target is caused by different tiles having the same $n_{\rm c}$ encoded bonds. Given the target fragments in the gray box, once components 6 and 8 are assembled, there are only 50\% chance to continue the yellow target. 
   {\bf B.} Likewise, components 4 and 1 have the same three encoded bonds and are thus indistinguishable for this assembly configuration.} 
\label{fig:error_prone}
\end{figure}

{\bf Retrieval with component reuse.}
Beyond conditions $\mu>0$ and $\delta>\delta_{\rm min}$, component reuse across $S>1$ targets further constrains assembly accuracy. We hereby discuss errors that arise from accelerated incorrect additions creating up to $n_{\rm c}$ bonds. Avoiding such combinatorial errors requires minimizing overlap among targets [Fig.~\ref{fig:error_prone}], which in turn limits the encoding capacity $S_{\rm max}$. 

\begin{figure}[!h] 
	\centering
    \includegraphics[width=.325\linewidth]{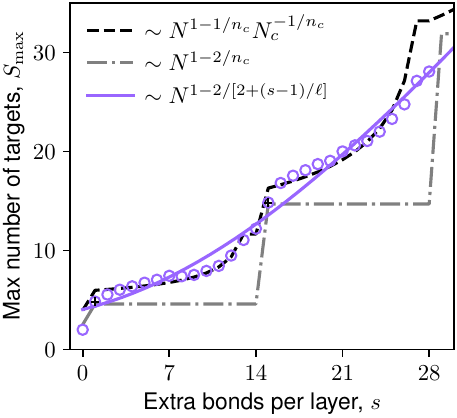}
   \caption{{\it Several approximations for the increase in encoding capacity.} Fits from Eq.~\eqref{eq:Smax_Nc} in dashed black, Eq.~\eqref{eq:Smax_gen} in gray mixed symbols, and Eq.~\eqref{eq:SmaxvZ} in purple. Here again, $\ell=14$, $\mu=20$ and $\delta=10$.}
\label{fig:Smaxvs}
\end{figure}

To obtain $S_{\rm max}$, we approximate $K_{\rm C}$ as in Eq.~\eqref{eq:K_C} and $K_{\rm I}\approx \Omega_{\rm I,c}\exp(n_{\rm c}\delta)$, with multiplicity $\Omega_{\rm I,c}\le\Omega_{\rm I}$. We moreover estimate $\Omega_{\rm I,c}\approx\Omega_{\rm C} N_{\rm I}$, which combines the location multiplicity of correct additions $\Omega_{\rm C}\le \ell$ with a species multiplicity of incorrect additions $N_{\rm I}\le N-1$ depending on the number of encoded structures $S$. Altogether, this amounts to $K_{\rm I}\approx N_{\rm I}K_{\rm C}$, giving a rate of combinatorial errors
\begin{equation}\label{eq:p_err_NI}
	p_{\rm err}\approx\frac{N_{\rm I}}{1+N_{\rm I}} \quad,
\end{equation} 
independent of $\delta$ and target geometry. We thus recall results from the polymerization study~\cite{Benoist25_SA}. For $n_{\rm c}=1$, most components have distinct neighbors across targets, yielding $N_{\rm I}\approx S-1$, $p_{\rm err}\approx (S-1)/S$, and thus $S_{\rm max}=1$. For $n_{\rm c}\ge2$, the average number of confounding monomers sharing $n_{\rm c}$ bonds with the correct one scales as
\begin{equation}\label{eq:NI}
    N_{\rm I}\sim (S-1)\left(\frac{S-1}N\right)^{n_{\rm c}-1}\quad,
\end{equation}
which decreases with increasing $N$. Approximating the error rate as $p_{\rm err}\approx N_{\rm I}$, the low-error condition $N_{\rm c}p_{\rm err}\ll1$ gives
\begin{equation}\label{eq:Smax_Nc}
	S_{\rm max}\sim N^{1-1/n_{\rm c}}N_{\rm c}^{-1/n_{\rm c}} \quad. 
\end{equation}
With $n_{\rm c}$ defined by Eq.~\eqref{eq:nmin_def}, this estimate for the encoding capacity reproduces, at least qualitatively, the oscillations of period $\ell$ visible in Fig.~\ref{fig:Smaxvs}. If one further approximates the number of critical events as $N_{\rm c}\sim N$, one recovers the simpler combinatorial scaling given in Eq.~\eqref{eq:Smax_gen}. A further simplification is obtained by replacing the integer-valued number of bonds involved in critical events by the smooth approximation
\begin{equation}\label{eq:nmin_app}
	n_{\rm c}\approx 2+(s-1)/\ell\approx z/2 \quad
\end{equation} 
Substituting this into Eq.~\eqref{eq:Smax_Nc} yields the compact scaling form $S_{\rm max}\sim N^{\alpha(z)}$, with $\alpha\approx1-2/n_{\rm c}\approx1-4/z$, as given in Eq.~\eqref{eq:SmaxvZ}. This approximation captures well the overall trend observed in Fig.~\ref{fig:Z6+_and_allZ}G.


\begin{figure}[!h]
    \includegraphics[width=.5\linewidth]{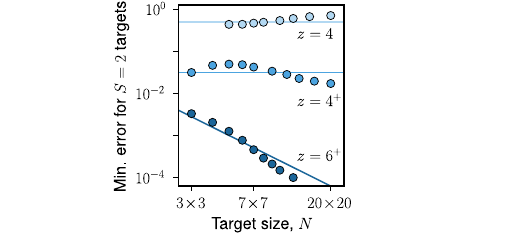}
   \caption{{\it Increasing connectivity reduces combinatorial errors.} Dependence of the minimum error fraction $e_{\rm min}$ at large $\delta$ on the target size for low promiscuity, \textit{i.e.} $S=2$. For $z=4$, $e_{\rm min}$ is near 50\%. For $z=4^+$ and $n_{\rm c}=2$, the error fraction drops below $5\%$ and varies little with target size [Eq.~\eqref{eq:error}]. For $z=6^+$ and $n_{\rm c}=3$, $e_{\rm min}$ drops again and follow the $1/N$ prediction (line).} 
\label{fig:error}
\end{figure}

Such combinatorial encoding results from a drastic reduction in combinatorial errors as $n_{\rm c}$ increases, most visible far below the encoding capacity $S\ll S_{\rm max}$. In that case, we can estimate the minimum error fraction in the large $\delta$ limit by expanding the accuracy with respect to the seeded target as follows:
\begin{equation}\label{eq:acc}
	1-e_{\rm min} = (1-p_{\rm err})^{N_{\rm c}} + \order{N_{\rm c}p_{\rm err}} \quad.
\end{equation} 
The first term on the right-hand side gives $P_{\rm ret}$ [Eq.~\eqref{eq:Pret_gen}], while the second term evaluates the dominant contribution to accuracy in the case of a few mistakes. 
In the low-error regime, $N_{\rm c}p_{\rm err}\ll1$, $e_{\rm min}$ scales as $\order{N_{\rm c}p_{\rm err}}$. For example, at $S=2\ll S_{\rm max}$, the minimum error fraction scales as $N_{\rm c}/N^{n_{\rm c}-1}$, as in Eq.~\eqref{eq:error}; see Fig.~\ref{fig:error}.

\section{Pathway selection}\label{sec.SI-path} 
We briefly noted in Sec.~\ref{sec.sol} that enhancing local connectivity at assembly bottlenecks improves performance by reducing the retrieval time $\tau_{\rm ret}$ and increasing the encoding capacity $S_{\rm max}$, primarily through a reduction of pathway multiplicity via a smaller effective $\Omega_{\rm C}$. We now make this connection explicit.

In the irreversible high-discrimination regime, the assembly of a single square structure of size $N=\ell\times\ell$ proceeds through a sequence of kinetically preferred steps. Starting from a small seed, the structure grows via $\order{N}$ successive additions, each occurring at a boundary site that maximizes the addition rate, \textit{i.e.} satisfying Eq.~\eqref{eq:it,xt}. At any given time $t$, the number of correct additions given the current boundary configuration is denoted $\Omega_{\rm C}^t$. The total pathway multiplicity is then given by
\begin{equation}\label{eq:P}
	P=\prod_{t=1}^{N}\Omega_{\rm C}^t\quad,
\end{equation}
which quantifies the combinatorial diversity of possible assembly routes compatible with correct growth. To understand how geometric and kinetic constraints influence retrieval accuracy and speed, we evaluate $P$, the retrieval time $\tau_{\rm ret}$, and the encoding capacity $S_{\rm max}$ under three representative scenarios of increasing structural constraint:

\begin{itemize}
\item {\bf Random boundary growth.} In the absence of bonding constraints, \textit{i.e.} for connectivity $z=0$, any of the $\ell$ boundary sites can accept any monomer species. The number of available additions therefore reads $\Omega_{\rm C}^t=\ell N$, giving an overall pathway multiplicity $P\approx(\ell N)^N\approx(\ell^3)^{\ell^2}$. In this limit, all growth paths are equally likely, retrieval is effectively random. At the same time, the absence of encoded bonds prevents acceleration of addition events via $\delta$, yielding $\tau_{\rm ret}\sim\exp(0)$, and the system can reliably encode only a single structure, \textit{i.e.} $S_{\rm max}=1$.

\item {\bf Nearest-neighbor case.} When bonds restrict which components can attach, the boundary configurations become more specific, as illustrated in Fig.~\ref{fig:Z4}D. For connectivity $z=4$, the number of possible additions is thereby reduced to $\Omega_{\rm C}^t=2$, except at the $\ell$ layer-nucleation events where $\Omega_{\rm C}^t=\ell$ choices remain, and at the $\ell$ layer-closing events where $\Omega_{\rm C}^t=1$. The resulting pathway multiplicity scales as $P\approx\ell^\ell 2^{N-2\ell}\approx\ell^\ell 2^{\ell(\ell-2)}$, leading to faster retrieval, $\tau_{\rm ret}\sim\exp(-\delta)$, while the encoding capacity remains limited to $S_{\rm max}=1$.

\item {\bf Funnel design.} Introducing a small number of additional bonds, such that $z=4^+$ [Fig.~\ref{fig:Z4+}A], further constrains assembly, channeling growth along a smaller subset of highly favorable paths. In this ``funneled'' regime, the multiplicity at layer-nucleation events reduces to $\Omega_{\rm C}^t=1$, and $P$ decreases to $P\approx2^{N-2\ell}\approx 2^{\ell(\ell-2)}$. This narrowing of viable pathways increases the kinetic bias toward correct assembly, yielding faster retrieval, $\tau_{\rm ret}\sim\exp(-2\delta)$, and a markedly higher encoding capacity, $S_{\rm max}\approx5$.
\end{itemize}

Together, these results demonstrate that kinetic encoding can be enhanced by shaping the assembly landscape: reducing pathway multiplicity through targeted connectivity increases effectively selects faster, more accurate assembly routes.

\section{Variation in discrimination strength and chemical potential}\label{sec.SI-Var}
In the main text, we assumed a uniform discrimination barrier $\delta$ for all encoded bonds and a uniform chemical potential $\mu$ for all component species. This is clearly a simplifying assumption, as in the cellular environment we expect a large variability in these parameters between species, $i=1,\dots,N$. To explore
the role of such variability, we sampled the chemical potential of each species $\mu_i$ and the discrimination barriers between two component species that are neighbors in a given complex $\delta_{ij}=\delta_{ji}$ from two Normal distributions $\mathcal N(\bar\mu,\sigma_\mu^2)$ and $\mathcal N(\bar\delta,\sigma_\delta^2)$, respectively. Our claim that variations in discrimination strengths or chemical potentials do not qualitatively alter the results is based both on theoretical arguments and on additional numerical simulations which we present below.

\begin{figure}[!t] 
   \centering
   \includegraphics[width=.35\linewidth]{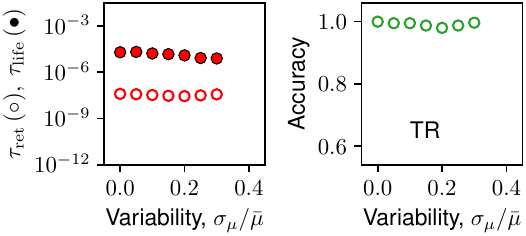}\\ \vspace{2mm}
   \includegraphics[width=.35\linewidth]{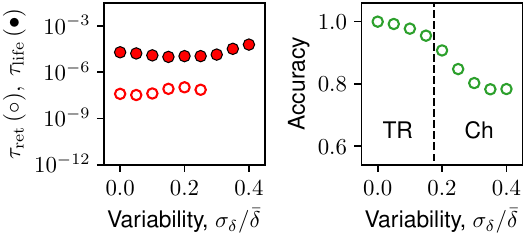}\\
\vspace{2mm}
   \includegraphics[width=.35\linewidth]{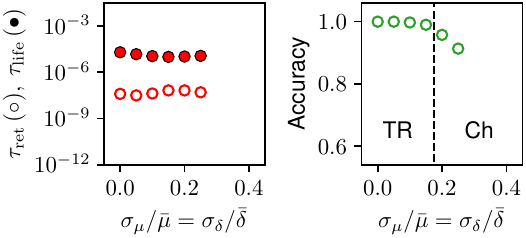}
   \caption{{\it Effect of chemical potential and discrimination energy variability in the transient retrieval regime.} Simulation outcomes for growth from a two-layer seed in systems with connectivity $z=4+2/\ell$ and $S=2$ encoded structures of side length $\ell=7$. The parameters $\mu$ and $\delta$ were independently sampled from the normal distributions $\mathcal N(\bar\mu,\sigma_\mu^2)$ and $\mathcal N(\bar\delta,\sigma_\delta^2)$, with means fixed in the transient retrieval (TR) regime at $\bar\mu=20$ and $\bar\delta=10$ [Fig.~\ref{fig:diags_varZ}A]. Increasing variability in the discrimination energy eventually drives a transition to the chimeric regime (Ch), as reflected by the decline in retrieval accuracy. By contrast, the retrieval time and lifetime remain only weakly affected over the range of variability considered. Large variability in the chemical potential instead promotes repeated addition and removal events, leading to long-lived fluctuations, as illustrated in Fig.~\ref{fig:Z4-SI}C.}
\label{fig:mu_delta_var}
\end{figure}

For $\sigma_\delta/\bar\delta$ and $\sigma_\mu/\bar\mu$ small, we expect that our results for $\sigma_\delta=\sigma_\mu=0$ still hold. However, differences emerge once the fluctuations become comparable to the mean values, particularly in the far-from-equilibrium regime $\mu\to\infty$, where discrimination barriers control the dynamics. 
To see this, consider a growing structure containing components $i$ and $j$, and two competing incoming components $k$ and $l$. Suppose $k$ can bind only to $i$, with discrimination barrier $\delta_{ik}$, whereas $l$ can bind simultaneously to both $i$ and $j$, with total barrier $\delta_{il}+\delta_{jl}$. The analysis developed throughout this work relies on the typical inequality
\begin{equation}
	\delta_{il}+\delta_{jl}>\delta_{ik} \quad,
\end{equation}
namely that the fully matched addition is kinetically favored over the partially matched one. When the fluctuations in $\delta_{ij}$ become of the same order as the mean, $\sigma_\delta\sim\bar\delta$, this ordering is no longer guaranteed: some partially matched additions may then proceed faster than fully matched ones. The kinetic hierarchy underlying accurate retrieval is therefore progressively lost. As a result, increasing variability in the discrimination barriers is expected to drive a crossover from a high-accuracy retrieval regime to a lower-accuracy chimeric regime. This expectation is borne out by the numerical results shown in Fig.~\ref{fig:mu_delta_var}.
\\

\newpage

\section{Extended discussion}\label{sec.SI-disc}

	\subsection{A toy model for ribosome self-assembly} 
We study a toy model for the self-assembly of large heteromeric protein complexes, such as ribosomes. Our framework relies on a variety of simplifying assumptions. In the following, we detail the justification regarding the main assumptions and discuss the potential effects of relaxing some of them.\\

{\bf No diffusion.}
We have omitted notions of component diffusion in the definition of our rates [Eq.~\eqref{eq:rates}]. In the context of cellular self-assembly, abundances of proteins vary wildly, ranging from only a few copies to millions per cell~\citep{Beck11}. While in certain cases self-assembly may be diffusion-limited, in many others it is not. As a particular example relevant to our work, ribosomal proteins are densely concentrated in the nucleolus, where ribosome assembly initially happens~\citep{Lafontaine21}. This is evidenced by the high difference in intensity between the nucleoli and the nucleoplasm. Growing mammalian cells synthesize $10^7$ new ribosomes at each division~\citep{Alberts15}, and hence produce $10^4$ ribosomes per minute~\citep{Lewis00}. These high production rates are more consistent with catalytic acceleration of ribosome assembly, than with diffusion-limited self-assembly. 
\\

{\bf No depletion.}
Parallel self-assembly of multiple copies of a structure introduces competition between assemblies for shared protein components~\cite{Murugan_Nat15}. The strength of this competition depends on the balance between the rates of component production and complex assembly. In the context of ribosome assembly, for example, ribosomal proteins are synthesized in the cytoplasm and imported into the nucleolus during ribosome assembly~\citep{Alberts15}, so the extent of stoichiometric depletion depends on the relative kinetics of component production and assembly. Such depletion is neglected in our model, which assumes an effectively infinite reservoir of monomers in order to isolate the consequences of kinetic encoding. This infinite-reservoir regime corresponds to the limiting case in which component production is sufficiently rapid to replenish the pool of free proteins on the timescale of assembly.
\\ 

{\bf Boundary growth.}
For simplicity, the manuscript focuses on quasi-2D boundary growth. In more general self-assembly simulations, binding between components outside the growing structure is often allowed. Incorporating such pathways can qualitatively alter the assembly dynamics, particularly far from equilibrium, where simulations may produce structures with holes or extended fiber-like growth that are more difficult to characterize systematically. 
While we have previously experimented with such simulations, the present work adopts the quasi-2D framework introduced by Whitelam and others~\citep{Whitelam14,Nguyen16,Ravasio24}, which yields cleaner and more controlled numerical results. Extending the model to fully 2D growth, including cluster diffusion and off-boundary binding events, remains an important direction for future work.
\\

{\bf Two-dimensional lattice framework.} 
While we have focused on a quasi-2D growth framework, we expect that in 2D and 3D, regardless of the particular geometry, kinetic discrimination involves a hierarchy of timescales and that minimal intervention is maximally effective at bottlenecks. We therefore expect many of our analytical results, including our expressions for $\delta_{\rm min},\tau_{\rm ret},e_{\rm min}$ in the main text and Eq.~~\eqref{eq:Smax_Nc} for $S_{\rm max}$, to extend to other geometries. 

These expressions are already sufficiently general to apply in 1D, provided the appropriate forms of $n_{\rm c},N_{\rm c},\Omega_{\rm I}$ and $\Omega_{\rm C}$ are used. Indeed, to bridge with the results of our previous polymerization study~\citep{Benoist25_SA}, we simply set $n_{\rm c}=z/2$, $N_{\rm c}=N$, $\Omega_{\rm I}=N-1$ and $\Omega_{\rm C}=1$. This consistency supports the expectation that the expressions for $\delta_{\rm min},\tau_{\rm ret},e_{\rm min}$ and $S_{\rm max}$ remain valid across different geometries and dimensions. Nonetheless, our lattice-based framework constitutes a strong simplification relative to the structural heterogeneity of, \textit{e.g.}, ribosomes.

	\subsection{Enhanced kinetic discrimination via allosteric interactions} 
\begin{figure}[!h] 
   \centering
   \includegraphics[width=.75\linewidth]{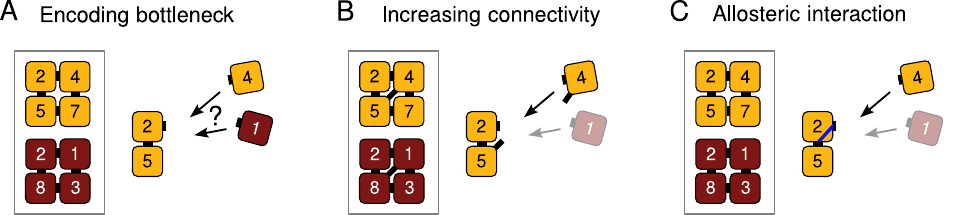}
   \caption{{\it Allosteric interactions can alternatively enhance kinetic discrimination.} {\bf A.} Given the target fragments in the gray box, kinetic discrimination between components $4$ and $1$ is impossible due to the single bond created. {\bf B.} In the main text, we resolve this encoding issue by adding diagonal bonds. {\bf C.} Allosteric interactions can likewise improve kinetic discrimination by allowing binding at one side of a monomer to influence binding at the other sides. That is, binding between components $2$ and $4$ is influenced by the bond between components $2$ and $5$ (blue link). This can promote continuation of the yellow structure, rather than switching to the red structure.}
\label{fig:allostery_drawing}
\end{figure}

In our initial framework with bulk connectivity $z=4$, bottlenecks arise from the presence of a single binding partner at the onset of a new layer, \textit{i.e.} $n_{\rm c}=1$, which renders binding both slower and more error-prone; see Figs.~\ref{fig:Z4_S=2}D and~\ref{fig:allostery_drawing}A. Rather than increasing the connectivity of components involved in these bottlenecks [Fig.~\ref{fig:allostery_drawing}B], one could alternatively introduce allosteric interactions that enhance kinetic discrimination through this single binding partner, leveraging its correct bonds with neighboring components in the growing structure; see Fig.~\ref{fig:allostery_drawing}C. 

To illustrate, consider a growing structure in which boundary components $2$ and $5$ match according to the yellow structure, and two incoming components can bind to component $2$. Component $4$ is compatible with the yellow structure, whereas component $1$ corresponds to the red structure. An allosteric mechanism could be invoked whereby the bond between components $2$ and $5$ modulates the interaction of component $2$ with incoming monomers, preferentially stabilizing bindings consistent with the yellow target. A detailed implementation of such allosteric effects lies beyond the scope of the present work.
\\

\end{document}